\documentclass[aps,pra,reprint,superscriptaddress,longbibliography]{revtex4-1}
\usepackage{amsmath}
\usepackage{graphicx}
\usepackage{amsfonts}
\usepackage{color}
\usepackage{bm}
\usepackage{lipsum}
\usepackage{comment}

\usepackage{amsmath}
\usepackage{amssymb}
\usepackage{graphicx} 
\usepackage{braket}
\usepackage{mathtools}
\usepackage{times}
\usepackage{enumerate}

\newcommand{\ii}{\mathrm{i}}

\newcommand{\Tr}{\mathrm{Tr}}

\newcommand*{\defeq}{\mathrel{\vcenter{\baselineskip0.5ex\lineskiplimit0pt\hbox{\scriptsize.}\hbox{\scriptsize.}}}=}

\usepackage[colorlinks,linkcolor=blue,urlcolor=blue,citecolor=blue]{hyperref}

\begin{document}

\title{Bath-induced interactions and transient dynamics in open quantum systems at strong coupling: Effective Hamiltonian approach}

\author{Marlon Brenes}
\email{marlon.brenes@utoronto.ca}
\affiliation{Department of Physics and Centre for Quantum Information and Quantum Control, University of Toronto, 60 Saint George St., Toronto, Ontario, M5S 1A7, Canada}

\author{Brett Min}
\affiliation{Department of Physics and Centre for Quantum Information and Quantum Control, University of Toronto, 60 Saint George St., Toronto, Ontario, M5S 1A7, Canada}

\author{Nicholas Anto-Sztrikacs}
\affiliation{Department of Physics and Centre for Quantum Information and Quantum Control, University of Toronto, 60 Saint George St., Toronto, Ontario, M5S 1A7, Canada}

\author{Nir Bar-Gill}
\affiliation{Department of Applied Physics, The Hebrew University of Jerusalem, Jerusalem, Israel}
\affiliation{The Racah Institute of Physics, The Hebrew University of Jerusalem, Jerusalem, Israel}

\author{Dvira Segal}
\affiliation{Department of Physics and Centre for Quantum Information and Quantum Control, University of Toronto, 60 Saint George St., Toronto, Ontario, M5S 1A7, Canada}
\affiliation{Department of Chemistry, University of Toronto, 80 Saint George St., Toronto, Ontario, M5S 3H6, Canada}

\begin{abstract}
Understanding the dynamics of dissipative quantum systems, particularly beyond the weak coupling approximation, is central to various quantum applications. While numerically exact methods provide accurate solutions, they often lack the analytical insight provided by theoretical approaches. 
In this study, we employ the recently-developed method dubbed the \emph{effective Hamiltonian theory} to understand the dynamics of system-bath configurations without resorting to a perturbative description of the system-bath coupling energy. Through a combination of mapping steps and truncation, the effective Hamiltonian theory offers both analytical insights into signatures of strong couplings in open quantum systems and a straightforward path for numerical simulations.
To validate the accuracy of the method, we apply it to two canonical models: a single spin immersed in a bosonic bath and two noninteracting spins in a common bath. In both cases, we study the transient regime and the steady state limit at nonzero temperature, and spanning system-bath interactions from the weak to the strong regime.
By comparing the results of the effective Hamiltonian theory with numerically exact simulations, we show that although the former overlooks non-Markovian features in the transient equilibration dynamics, it correctly captures non-perturbative bath-generated couplings between otherwise non-interacting spins as observed in their synchronization dynamics and correlations. Altogether, the effective Hamiltonian theory offers a powerful approach to understanding strong coupling dynamics and thermodynamics, capturing the signatures of such interactions in both relaxation dynamics and in the steady state limit.  
\end{abstract}

\date{\today}

\maketitle

\section{Introduction}

The progress in experimental platforms, 
such as cold atoms~\cite{Kinoshita:2006,Trotzky:2012,Kaufman:2016,Bernien:2017,Tang:2018,Jepsen:2020} 
and nitrogen-vacancy centers in diamond~\cite{Maze:2008,BarGill12}, has enabled fundamental studies of 
quantum systems with coherence periods that can be maintained over extended timescales. 
However, even with the remarkable control achieved in such systems, 
it is only in specific scenarios that one can ignore the environmental effects on a generic quantum system. 
Indeed, in emerging quantum technologies with applications in, e.g., computation, the \emph{noise} is an ever-present effect that is considered a hindrance towards efficient implementations~\cite{Preskill:2018}.

There exists, in turn, a present interest in understanding environmental effects in quantum systems. Such is the case in, for instance, biological systems and light-harvesting complexes~\cite{Duan:2022}. 
Emerging technologies such as 
thermal machines~\cite{Kosloff:2014Rev,Goold:2016,Benenti:2017,Mitchison:2019,Linke:2018,Ronzani:2018,Mosso:2019,Carrega:2023},
which operate in the quantum regime, also require a deep understanding of system-environment effects for efficient operation.
Moreover, light-matter interactions between the vacuum fluctuations and a material placed inside the cavity provide a new scheme in controlling/engineering material/molecular properties~\cite{Francisco_2021, Hannes_2021,Owens_2022,Schlawin_2022,Bloch_2022, Galego_2015,Herrera_2016}.
An important remark is that in most of these cases, the system-environment interaction energy cannot be regarded as 
an asymptotically weak parameter in the configuration. 
This limits the applicability of standard perturbative open quantum systems theory~\cite{BreuerPetruccione}.

This brings us to the topic of strongly coupled open quantum systems~\cite{Strasberg:2019}. In the long time limit,
a quantum system weakly coupled to a thermal bath will generically equilibrate with its environment to a 
canonical thermal Gibbs state at the inverse temperature $\beta$~\cite{BreuerPetruccione}. 
On the other hand, in situations where the interaction energy between the system and the bath is the largest energy scale 
in the configuration, such equilibration occurs to a \emph{noncanonical} quantum state, 
which inherits the microscopic properties of the system-bath interaction~\cite{Cresser:2021Ultrastrong}. 
The intermediate regime is intricate and also leads to non-canonical equilibrium states, 
which can be understood microscopically for specific spectral density functions that dictate the system-bath interaction~\cite{Segal:2023Effective}.  

Starting from out-of-equilibrium initial conditions, it is also of interest to understand 
the {\it transient} dynamics that lead the system and bath to the state of thermal equilibrium. 
In this respect, many impressive techniques have come to fruition in the last three decades to target the problem. 
These include; for example, path integral methods~\cite{Makri:1995,SegalPI10,Kilgour:2019,Kundu:2023,Strathearn:2018,TEMPO2}, 
hierarchical equations of motion (HEOM)~\cite{Tanimura:2020HEOM,Lambert:2023Qutip}, 
Monte Carlo approaches~\cite{Erpenbeck:2023,Cohen:2015,Cohen:2015_2}, 
techniques based on machine learning calculations~\cite{Luo:2022}, 
tensor networks~\cite{Prior:2010,Strathearn:2018,Brenes:2020TNT,Gauger22},
generalized quantum master equations \cite{GQME1a,GQME1b,GQME3}, Davydov ansatze~\cite{Lipeng:2017,Zhao:2023,Zhou:2016}
and diverse Markovian embeddings~\cite{Woods:2014Embedding,Chin:2010,Imamoglu:1994,Garraway1997a,Garraway1997b,Menczel:2024}; all of which can be applied efficiently to study dynamics of system-bath configurations depending on the nature of the microscopic model, energy scales, and computational costs.    

These are most impressive developments, yet given their numerical nature, 
these tools are limited in the basic understanding they can offer on the nature of strong coupling phenomena. In this work, we take a step forward in the direction of employing a semi-analytical approach
towards understanding transient dynamics of out-of-equilibrium system-bath configurations with experimental relevance.  

The effective Hamiltonian theory~\cite{Segal:2023Effective,Nick:2023Effective} 
has been presented as an alternative to fully numerical techniques. Previous works, focused on the equilibrium and steady-state behavior of impurity systems coupled to bosonic thermal reservoirs, showed that this theory can provide an understanding of strong coupling effects
in thermal and electronic devices
without assumptions on the energy scale of the system-bath interaction energy. 
The theory relies on a Markovian embedding with a subsequent unitary transformation that imprints environmental effects onto the system's degrees of freedom. One then considers a specific sub-manifold of the resulting Hamiltonian, and such an approximation can be justified under certain assumptions on the temperature and the bath's spectral function.
As a whole, the effective Hamiltonian mapping prepares a dressed system Hamiltonian that embeds strong coupling effects. The effective system itself is assumed to couple to a residual bath in a Born-Markov limit~\cite{BreuerPetruccione}.
With this method, one can obtain analytical expressions for equilibrium states in the weak, intermediate, and strong system-bath coupling regimes~\cite{Nick:2023Effective}. Furthermore, the technique can be employed even for complex quantum systems that may display phase transitions induced by collective or individual baths~\cite{Min:2023}. 

Here, we assess the capability of the effective Hamiltonian theory to capture and elucidate
both the equilibrium and transient dynamics of system-bath configurations that are initially prepared out of equilibrium. Although this task can be addressed from e.g., numerical renormalization group (NRG) techniques~\cite{Orth:2010} and multi-D$_1$ ansatze~\cite{Deng:2016TwoSpin} at zero temperature with semi-analytical arguments, here we focus on nonzero temperature baths. 

We test the effective Hamiltonian method on two prototype models of systems composed of spins: 
A spin-boson model, with a single spin impurity immersed in a boson bath, and the scenario with 
two noninteracting spins embedded in a common bath. In both cases, we examine the relaxation dynamics and the steady state limit while exploring system-bath coupling energy from the regime of weak but finite coupling to the strong coupling limit. 

We find that the theory can capture certain strong coupling effects in the dynamics, but it falls short in capturing other features, specifically those pertaining to non-Markovian effects: For a single dissipative spin, the effective Hamiltonian method reproduces the correct relaxation timescale, 
but it fails to account for secondary non-Markovian features that show as oscillatory relaxation dynamics. In the case of two spins in a bath, the theory depicts the nontrivial and nonperturbative effects of bath-induced spin-spin interactions: It reproduces the correct synchronization oscillation frequency, as well as the generation of correlations between spins in the transient domain and their behavior in steady state.
The method however tends to overestimate the magnitude of spin oscillations, and it overlooks additional fine details in the dissipative dynamics.

On a more general footing, we find that the effect of a common bath on a quantum system comprising individual spins is to generate interactions between the spins and to build correlations among them, even when the bath is maintained at nonzero temperature. These interactions result in synchronized dynamics of the spins, resembling that observed in atomic ensembles under the effect of a single cavity mode~\cite{Xu:2014Sync}. Such synchronized dynamics has been discussed in specific models, e.g., in Refs.~\cite{Giorgi:2013Sync,Cabot:2019Sync,Karpat:2020Sync,Xiao:2023Sync}. Here, we understand this effect and the relevant energy scales {\it analytically} from the effective Hamiltonian theory.

As for the equilibrium (long-time) behavior, the effective Hamiltonian method qualitatively captures the polarization dynamics from the weak to the ultrastrong coupling limit in the spin-boson model. Notably, it attains exact results in both the ultraweak and ultrastrong coupling limits as shown in Refs.~\cite{Segal:2023Effective,Nick:2023Effective}.
Interestingly, we show with simulations that in the steady state limit the effective Hamiltonian method achieves not only qualitative but {\it quantitative} accuracy for scenarios involving two or three spins immersed in a common bath, across the entire coupling range. Such a trend suggests that the effective Hamiltonian method could become progressively more reliable as the complexity of the system, as measured by the number of spins, grows. 

In Sec.~\ref{sec:main} we describe our models, the theory of effective Hamiltonian, and the Redfield quantum master equation (QME) employed to describe the transient dynamics. Our results are divided into Sec.~\ref{sec:thermaleq} and Sec.~\ref{sec:dynamics_results}, devoted to equilibrium properties and dynamical signals, respectively. In light of our results, we provide a general assessment of the effective Hamiltonian method for both transient or steady-state problems in Sec.~\ref{sec:assess}. Concluding remarks and the experimental relevance of our results are included in Sec.~\ref{sec:conclusion}.

\section{Method and bath-induced interactions}
\label{sec:main}

This Section is devoted to the description of the models, methods and techniques employed in our work. In Sec.~\ref{sec:model}, we describe the open quantum system model we consider here. 
In Sec.~\ref{sec:effective}, we present the analytical effective Hamiltonian treatment used to study dynamics. 
The simulation method we use is explained in Sec.~\ref{sec:dynamics}. We also perform a spectral analysis of the effective Hamiltonian in Sec. \ref{sec:eigen}. This analysis is relevant for understanding the dynamics of out-of-equilibrium configurations. 

\subsection{Spin-bath models}
\label{sec:model}

We shall consider the out-of-equilibrium dynamics of a quantum system coupled to a bosonic bath. The total Hamiltonian is described by ($\hbar = 1$)
\begin{align}
\label{eq:h_2spin_total}
\hat{H} =  \hat{H}_S + \hat{S}\sum_k t_k  \left( \hat{c}_k^{\dagger} + \hat{c}_k \right) + \sum_k \nu_k \hat{c}_k^{\dagger} \hat{c}_k,
\end{align}
where $\hat{H}_S$ is the system Hamiltonian. Here, $\hat{S}$ is an interaction operator with support over the system's degrees of freedom. It describes the action of the bath on the system, with displacement coupling on the bath degrees of freedom. The set $\{ \hat{c}_k \}$ are bosonic operators, and $\nu_k$ and $t_k$ are the frequencies and the environment-system couplings between the $k$-th bosonic mode in the reservoir and the system, respectively. 

\begin{figure}[t]
\fontsize{13}{10}\selectfont 
\centering
\includegraphics[width=0.7\columnwidth]{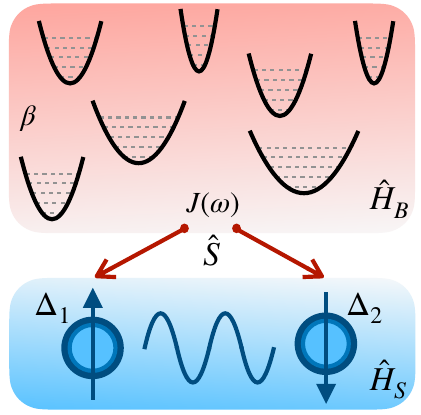}
\caption{A two-spin system ($\hat{H}_S$) with spin splittings $\Delta_1$ and $\Delta_2$ is coupled to a thermal bath ($\hat{H}_B$) at inverse temperature $\beta$ with a coupling operator $\hat{S}$, modeled by an infinite collection of harmonic modes. The spectral density of the reservoir is given by $J(\omega)$. The spins are not coupled to each other but an effective coupling arises from the interaction with the common bath.}
\label{fig:0}
\end{figure}

For the system itself, we consider a spin-1/2 Hamiltonian comprising $N$ spins of the form
\begin{align}
\label{eq:h_and_s_system}
\hat{H}_S = \sum_{i=1}^{N} \Delta_i \hat{\sigma}^z_i \quad {\rm and} \quad \hat{S} = \sum_{i=1}^{N} \hat{\sigma}^x_i,
\end{align}
where $\hat{\sigma}^{\alpha}_i$ is the Pauli spin operator in the $\alpha = \{x, y, z\}$ direction of the $i$-th spin. 
A diagrammatic depiction is shown in Fig.~\ref{fig:0}. Note that in our model, we assume that there are no direct interactions between each (spin) subsystem, though these interactions can be included~\cite{Min:2023}. However, as we show in Sec.~\ref{sec:effective}, an effective interaction emerges between these spins due to their finite coupling to the same bath. 
In light of this, we will address the problem in the strong system-bath coupling regime, in which conventional perturbative theories fail. 

The effect of the environment is typically described via a spectral density function, which may be defined as \mbox{$J(\omega) = \sum_k |t_k|^2 \delta(\omega - \nu_k)$}. The choice of $\hat{S}$ has intricate consequences for the dynamics of the system and, whenever the interaction energy between the system and environment is non-negligible, 
it will also affect the thermal equilibrium state~\cite{Segal:2023Effective}. 
In this work, we shall focus on the case whereby the coupling to the environment 
generates coherent evolution towards the equilibration of the system, by choosing $\hat{S}$ along a different component than that of the spin-system Hamiltonian. 

For the analysis of the dynamics, we focus on the $N = 1, 2$ models.
The $N=1$ scenario corresponds to the canonical spin-boson model. The $N=2$ case exemplifies the nontrivial physics of bath-induced coupling between spins.

\subsection{Reaction-coordinate mapping and effective Hamiltonian theory}
\label{sec:effective}

We want to make minimal assumptions about the system-bath interaction energy and, therefore, we cannot resort to the conventional weak-coupling open systems theory~\cite{BreuerPetruccione}. 
For certain spectral density functions of the bath, $J(\omega)$, it is beneficial to employ the reaction-coordinate mapping to go beyond 
the weak-coupling approximation~\cite{Hughes:2009RC,Hughes:2009RC2,Martinazzo:2011RC,Iles:2014RC,NazirJCP16,Strasberg:2016RC,Hughes:2009RC,Nick:2021RC, Luis19,Gernot19}. 

For the model at hand, we extract a single collective reaction coordinate (RC) from the bath to which the system is strongly coupled. 
This mapping is useful if the residual bath, i.e., the one composed of the degrees of freedom of the bath remaining after the extraction of the reaction coordinate, 
is weakly coupled to the enlarged system Hamiltonian. The enlarged system Hamiltonian is composed of the original system and the reaction coordinate. After such mapping, Eq.~\eqref{eq:h_2spin_total} translates to~\cite{Hughes:2009RC,Hughes:2009RC2,Martinazzo:2011RC,Iles:2014RC,NazirJCP16,Strasberg:2016RC,Nick:2021RC,Segal:2023Effective,Luis19,Gernot19}
\begin{align}
\label{eq:h_rc_tot_2}
\hat{H}^{\rm{RC}} &= \hat{H}_S + \Omega \hat{a}^{\dagger} \hat{a} + \lambda \hat{S} \left( \hat{a}^{\dagger} + \hat{a} \right) \nonumber \\
&+ \sum_k f_k \left( \hat{a}^{\dagger} + \hat{a} \right) \left( \hat{b}_k^{\dagger} + \hat{b}_k \right) + \sum_k \omega_k \hat{b}_k^{\dagger} \hat{b}_k.
\end{align}
In the above equation, the bosonic operators $\hat{a}$ and $\hat{a}^{\dagger}$ pertain to the reaction coordinate with frequency $\Omega$. 
The parameter $\lambda$ is the coupling between the original system $\hat{H}_S$ and the reaction coordinate via the system operator $\hat{S}$. 
The enlarged system, which is composed of $\hat{H}_S$ and the reaction coordinate, is now coupled to a residual bath with the effect described by the new spectral function $J_{\rm RC}(\omega) = \sum_k |f_k|^2 \delta(\omega - \omega_k)$. 
The harmonic modes of the residual reservoir are described via the sets $\{ \hat{b}_k \}$ with frequencies $\omega_k$, which are linear combinations of the original harmonic modes $\{ \hat{c}_k \}$. 
Both $\lambda$ and $\Omega$ may be obtained from the original spectral function 
through the relations $\lambda^2 = \frac{1}{\Omega} \int_0^\infty {\rm{d}}\omega \; \omega J(\omega)$ and $\Omega^2 = \frac{\int_0^\infty {\rm{d}}\omega \; \omega^3 J(\omega)}{\int_0^\infty {\rm{d}}\omega \; \omega J(\omega)}$~\cite{Iles:2014RC}.

In the case of a system comprising multiple spins as in Eq.~(\ref{eq:h_and_s_system}), we see from Eq.~(\ref{eq:h_rc_tot_2}) that the RC couples to all spins through $\hat S$, and thus it is anticipated that the spins would develop effective interactions between each other through the RC. However, estimating the magnitude of this interaction directly from Eq. (\ref{eq:h_rc_tot_2}) is not straightforward.
We next show that with an additional polaron mapping and a truncation, we can explicitly extract this bath-mediated spin-spin coupling.

An effective Hamiltonian treatment may be employed in this model, which has been successful at describing equilibrium thermodynamics~\cite{Segal:2023Effective,Nick:2023Effective}. We define the so-called polaron transformation
[see discussion below Eq.~\eqref{eq:h_eff_polaron}]
\begin{align}
\hat{U} = {\rm{exp}} \left[ \frac{\lambda}{\Omega} (\hat{a}^{\dagger} - \hat{a}) \hat{S} \right].
\label{eq:polaron}
\end{align}
The bosonic degrees of freedom can be found to transform to $\hat{U} \hat{a}^{(\dagger)} \hat{U}^{\dagger} = \hat{a}^{(\dagger)} - \frac{\lambda}{\Omega}\hat{S}$ following a Baker–Campbell–Hausdorff expansion~\cite{BreuerPetruccione}. 
After the mapping we get
\begin{align}
\hat{\tilde{H}} &= \hat{\tilde{H}}_S + \Omega \hat{a}^{\dagger} \hat{a} - \frac{\lambda^2}{\Omega}\hat{S}^2 \nonumber \\
&+ \sum_k f_k \left( \hat{a}^{\dagger} + \hat{a} - \frac{2\lambda}{\Omega} \hat{S} \right) \left( \hat{b}_k^{\dagger} + \hat{b}_k \right) + \sum_k \omega_k \hat{b}_k^{\dagger} \hat{b}_k,
\end{align}
where $\hat{\tilde{H}}_S = \hat{U} \hat{H}_S \hat{U}^{\dagger}$.
An effective Hamiltonian may be obtained when projecting the polaron-transformed Hamiltonian 
into the ground-state manifold of the reaction coordinate via the projector $\hat{Q}_0 = \ket{0} \bra{0}$, which yields
\begin{align}
\label{eq:h_eff_total}
\hat{H}^{\rm{eff}} &= \hat{Q}_0 \hat{\tilde{H}}_S \hat{Q}_0 - \frac{\lambda^2}{\Omega}\hat{S}^2 \nonumber \\
&- \hat{S}\sum_k \frac{2\lambda f_k}{\Omega}  \left( \hat{b}_k^{\dagger} + \hat{b}_k \right) + \sum_k \omega_k \hat{b}_k^{\dagger} \hat{b}_k.
\end{align}
It is important to remark that Eq.~\eqref{eq:h_eff_total} has the same form as 
Eq.~\eqref{eq:h_2spin_total}, with re-scaled coupling energy to the residual reservoir.
Importantly, strong coupling effects are embedded into the new effective system Hamiltonian, given by
\begin{align}
\label{eq:h_eff_2}
\hat{H}^{\rm{eff}}_S = \hat{Q}_0 \hat{\tilde{H}}_S \hat{Q}_0 - \frac{\lambda^2}{\Omega}\hat{S}^2.
\end{align}
An analytical expression may be obtained for Eq.~\eqref{eq:h_eff_2}. We show in Appendix~\ref{ap:eff_h} that
for the model in Eq.~\eqref{eq:h_and_s_system} the application of the polaron mapping and the truncation of the energy spectrum leads to
\begin{align}
\label{eq:h_eff_explicit}
\hat{H}^{\rm{eff}}_S = e^{-\frac{2\lambda^2}{\Omega^2}} \hat{H}_S - \frac{\lambda^2}{\Omega} \hat{S}^2.
\end{align}
Thus, the signatures of strong coupling within the effective system are twofold:
(i) The spin splittings $\Delta_i$ are exponentially-suppressed with an exponent 
$-2\lambda^2/\Omega^2$. (ii) An effective interaction, from the action of the environment, 
arises from the $\hat{S}^2$ term with the $\lambda^2/\Omega$ factor.

In particular, if $N = 2$ then Eq.~\eqref{eq:h_eff_explicit} translates to
\begin{align}
\label{eq:h_eff_2_explicit}
\hat{H}^{\rm{eff}}_S = e^{-\frac{2\lambda^2}{\Omega^2}} \left( \Delta_1 \hat{\sigma}^z_1 + \Delta_2 \hat{\sigma}^z_2 \right) - \frac{\lambda^2}{\Omega} \left( \hat{\sigma}^x_1 + \hat{\sigma}^x_2 \right)^2.
\end{align}
The above treatment reveals the structure of bath-induced interactions. Furthermore, we find from a polaron transformation applied to all bosonic models, directly onto 
Eq.~\eqref{eq:h_2spin_total} that, 
more generally, the form of the system Hamiltonian ($N = 2$) when one considers the effects of the bath is 
\begin{align}
\label{eq:h_eff_polaron}
\hat{H}_S^{\rm pol} = \tilde{\Delta}_1 \hat{\sigma}^z_1 + \tilde{\Delta}_2 \hat{\sigma}^z_2 - 2E_I \hat{\sigma}^x_1 \hat{\sigma}^x_2.
\end{align}
For details, see Appendix~\ref{ap:polaron}.
This suggests that, on a more general footing, the effect of the bath is to rescale the spin splitting parameters $\Delta_i$ and to introduce spin interactions, with interaction energy $E_I$. 
The rescaling of $\Delta_i$ and the form of the interaction naturally depends on the microscopic details of the model,
and they are formally given in Appendix~\ref{ap:polaron}.

The basic idea behind the polaron transformation is that the interaction between a quantum system (for example, electrons) and its environment (phonons) leads to the dressing of the quantum system by the environment, the former now referred to as a ``polaron".  In the case of a charged particle, the coupling to the lattice modes effectively increases the mass of the electron, thus slowing it down.

The polaron transformation, which is unitary, is designed  to (even if partially) decouple the quantum system from the environment. As an outcome, the system's Hamiltonian becomes dressed by the bath.
Studying quantum dynamics under a polaron picture is beneficial at strong system-bath coupling. This problem has been investigated in many works, with some examples in Refs.~\cite{Nazir:2009Polaron,Cao12,SBSegal1,SBSegal2}. The advantage of the polaron picture is that in the so-called polaron frame, one can identify a perturbative parameter that replaces the original (nonpertuabtive) system-bath coupling ~\cite{McCutcheon:2010}. 

In our approach, the transformation in Eq.~\eqref{eq:polaron} shifts the reaction-coordinate Hamiltonian into the so-called polaron frame. As showed in 
Ref.~\cite{Segal:2023Effective}, this step extends the range of applicability of the reaction-coordinate mapping.

A novel application of the effective Hamiltonian method, examined in this work, is that it exposes and thus allows us to extract, the energy scale of the spin-spin interaction, $E_I$, from dynamical response signals of local observables.
A complication arises due to the fact that the interaction with the bath results in two different effects: an emergent interaction energy between the spins (which we want to quantify) leading to Rabi-like oscillations, as well as the more standard relaxation dynamics to thermal equilibrium, which can show rich signatures when non-Markovian effects take place. 
Below, we show that it is possible to extract the interaction energy $E_I$ from dynamical signals for certain initial conditions. 

We also note that for the $N = 1$ case (the spin-boson model), $\hat{S}^2 = \mathbf{I}$ such that no new internal interactions emerge from our treatment of a single spin degree of freedom.

\subsection{Dynamics}
\label{sec:dynamics}

We study the dynamics of the system by employing four tiers of methods:

(i) The hierarchical equations of motion~\cite{Tanimura:2020HEOM,Lambert:2023Qutip}, 
a numerically-exact method. While the method is exact, for practical purposes in numerical calculations one needs to truncate the hierarchy of equations and the number of terms pertaining to the expansion of the bath correlation function. Due to the increased complexity that arises as the system-bath interaction energy increases, with this method we obtain accurate results in the weak-to-intermediate interaction energy regime. Details on the the HEOM are presented in Appendix~\ref{ap:heom}. 

(ii) The Redfield QME within the reaction coordinate mapping Hamiltonian. Our simulations (as well as previous works, see, e.g., Refs.~\cite{Iles:2014RC,NazirJCP16,Nick:2023Effective}) demonstrate that for a bath characterized by a Brownian spectral density function, this approach is accurate compared to HEOM simulations.

(iii) The Redfield QME on the effective Hamiltonian. Compared to approach (ii), 
this simulation offers a computational advantage since it involves a system Hamiltonian of the same Hilbert space dimension as the original problem, the result of the truncation of the reaction coordinate. Moreover, the analytical form of the effective Hamiltonian allows a rational interpretation of the results.

(iv) The Redfield QME used directly on the original Hamiltonian, thus assuming a weak coupling limit. In some cases, to illustrate the importance of adopting more accurate tools, (i)-(iii), we perform such simulations even beyond the weak coupling regime.
 
We provide next the details on the Redfield QME, focusing as an example on the {\it reaction-coordinate} Hamiltonian~\cite{Iles:2014RC,Nick:2021RC}, that is, case (ii) listed above.

The QME for the dynamics of the matrix elements of the density matrix $\hat{\rho}(t)$ reads 
\cite{NitzanBook}
\begin{align}
\label{eq:redfield_dyn}
\frac{{\rm d}\rho_{mn}}{{\rm d}t} &= -\ii \omega_{mn} \rho_{mn} \nonumber \\
&- \sum_{jk} \left[ R_{mj,jk}(\omega_{kj}) \rho_{kn} + R^*_{nk,kj}(\omega_{jk}) \rho_{mj} \right. \nonumber \\
&- \left. R_{kn,mj}(\omega_{jm}) \rho_{jk} - R^*_{jm,nk}(\omega_{kn}) \rho_{jk} \right],
\end{align}
where $\omega_m$ are eigenenergies of the reaction-coordinate system Hamiltonian, defined from 
\begin{align}
\label{eq:h_rc_system}
\hat{H}^{\rm RC}_S = \hat{H}_S + \Omega \hat{a}^{\dagger} \hat{a} + \lambda \hat{S} \left( \hat{a}^{\dagger} + \hat{a} \right),
\end{align}
such that $\hat{H}^{\rm RC}_S \ket{\omega_m} = \omega_m \ket{\omega_m}$. 
In Eq.~\eqref{eq:redfield_dyn}, the subindices denote the matrix elements in the energy eigenbasis while $R_{ab,cd}$ are the elements in this basis of the Redfield tensor. 
We have defined $\omega_{mn} = \omega_m - \omega_n$. The Redfield equation assumes that the dynamics begin from a system-bath factorized initial conditions.
In arriving to Eq.~\eqref{eq:redfield_dyn}, we note that the usual Born-Markov approximation has 
been carried out, but not the secular approximation.
To confirm that the dynamics generated in our model under the conditions we have presented are correct, specifically, the assumption of weak coupling and Markovianity to the residual bath,
we have compared the dynamics of observables computed from Eq.~\eqref{eq:redfield_dyn} with the numerically exact HEOM method.
In Appendix~\ref{ap:heom} we demonstrate that when solving the Redfield QME on the RC Hamiltonian by 
following Eq.~\eqref{eq:redfield_dyn} we get the same results as with the hierarchical equations of motion.

The Redfield tensor is evaluated from the bath correlation functions in frequency space, following the relation
\begin{align}
\label{eq:refield_tensor_integral}
R_{mn,jk}(\omega) &= S_{mn} S_{jk} \int_{0}^{\infty} {\rm d}\tau e^{\ii \omega \tau} \langle \hat{B}(\tau) \hat{B}(0) \rangle,
\end{align}
where $\hat{S} = \left( \hat{a}^{\dagger} + \hat{a} \right)$ is the system operator that couples to the bath, with matrix elements $S_{mn}$ $[S_{jk}]$ in the energy eigenbasis of $\hat{H}^{\rm RC}_S$. $\hat{B} = \sum_k f_k (\hat{b}^{\dagger}_k + \hat{b}_k)$
is the bath operator. 
The expectation value in Eq.~\eqref{eq:refield_tensor_integral} 
is taken with respect to the thermal state of the residual bath, i.e., $\langle \bullet \rangle = \Tr[\bullet \hat{\rho}_B]$, where $\hat{\rho}_B = e^{-\beta \hat{H}_B} / \Tr[e^{-\beta \hat{H}_B}]$, $\hat{H}_B = \sum_k \omega_k \hat{b}^{\dagger}_k \hat{b}_k$ and $\beta = 1 / T$ the inverse temperature. The integral can be readily evaluated to obtain
\begin{align}
\label{eq:refield_tensor_correlation}
R_{mn,jk}(\omega) &= S_{mn} S_{jk} \left[ \Gamma_{\rm RC}(\omega) + \ii \delta_{\rm RC}(\omega) \right],
\end{align}
where $\Gamma_{\rm RC}(\omega)$ is the symmetric part of the correlation function 
and $\delta_{\rm RC}(\omega)$ is the so-called Lamb shift term, which we neglect in our calculations \cite{Correa_Lamb}. 
We note that, in lack of a general solution for the diagonal form of 
$\hat{H}^{\rm RC}_S$, Eq.~\eqref{eq:redfield_dyn} is typically solved numerically with 
a truncated manifold of the reaction coordinate, 
such that only $M$ levels of the harmonic mode 
are kept to tractably study the dynamics.

To evaluate Eq.~\eqref{eq:refield_tensor_correlation}, one requires knowledge of the spectral density
function of the residual reservoir,  $J_{\rm RC}(\omega)$. This spectral function depends directly on the original spectral density, $J(\omega)$. 
We consider the specific case of a Brownian spectrum for the original spectral function
\begin{align}
\label{eq:spec_fun_brownian}
J(\omega) = \frac{4\gamma \Omega^2 \lambda^2 \omega}{(\omega^2 - \Omega^2)^2 + (2\pi\gamma\Omega\omega)^2},
\end{align}
for which the residual spectral function can be computed analytically and known to be Ohmic with an exponential cut-off~\cite{Iles:2014RC},
\begin{align}
\label{eq:spec_fun_rc_ohmic}
J_{\rm RC}(\omega) = \gamma \omega e^{-\omega / \Lambda},
\end{align}
where $\Lambda$ is a large energy cut-off and $\gamma$ is a dimensionless parameter capturing the coupling  to the residual reservoir.
With these considerations, one can evaluate Eq.~\eqref{eq:refield_tensor_correlation} with
\begin{align}
\Gamma_{\rm RC}(\omega) = \begin{cases} \pi J_{\rm RC}(\omega) n(\omega) & \omega > 0 \\ \pi J_{\rm RC}(|\omega|) [n(|\omega|) + 1] & \omega < 0 \\ \pi \gamma / \beta & \omega = 0, \end{cases}
\end{align}
where $n(\omega) = (e^{\beta \omega} - 1)^{-1}$ is the Bose-Einstein distribution. 

The dynamics generated by the bath from the perspective of the effective Hamiltonian theory follows the same steps as above, with the only difference being that, as opposed 
to considering the eigenbasis of $\hat{H}^{\rm RC}_S$ for the Redfield tensor, 
one uses the eigenbasis of the effective Hamiltonian $\hat{H}^{\rm eff}_S$ from Eq.~\eqref{eq:h_eff_2_explicit} along with the coupling operator $\hat{S} = \sum_{i=1}^{N} \hat{\sigma}^x_i$ for our particular choice in the spin model.

Similarly, the simulation of the system dynamics at tier (iv), with the original Hamiltonian, 
proceeds based on Eq.~(\ref{eq:h_and_s_system}) with the original spectral function $J(\omega)$.

\subsection{Eigenenergies of the effective Hamiltonian}
\label{sec:eigen}

\begin{figure}[t]
\fontsize{13}{10}\selectfont 
\centering
\includegraphics[width=1\columnwidth]{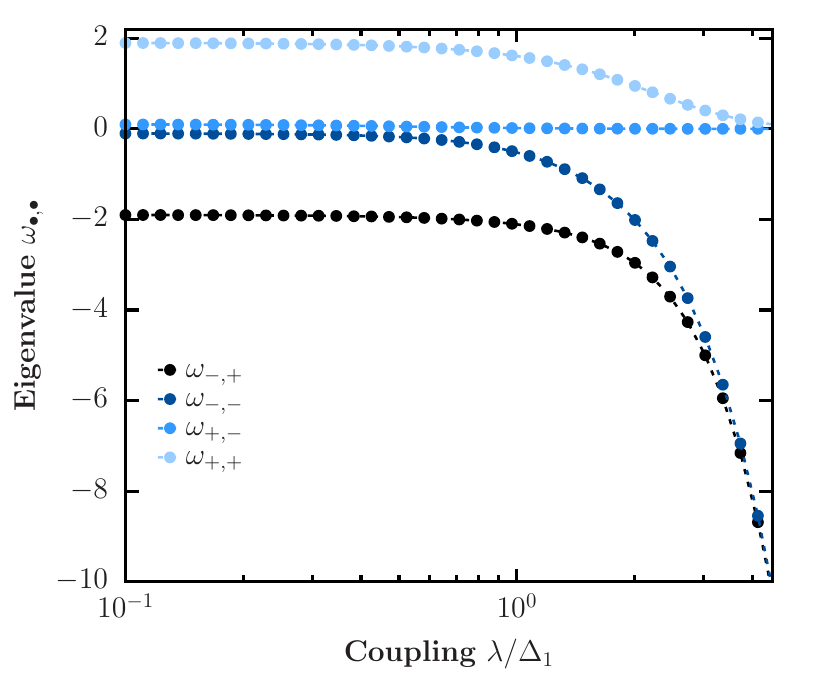}
\caption{Eigenvalues $\omega_{\mp, \mp}$ of the effective system Hamiltonian $\hat{H}^{\rm eff}_S$ ($N=2$) as a function of the coupling energy $\lambda$. Results are shown for $\Omega = 8\Delta_1$ and $\Delta_2 = 0.9\Delta_1$.}
\label{fig:1}
\end{figure}

We have shown that part of the effect of the bath becomes embedded into the effective Hamiltonian 
$\hat{H}^{\rm eff}_S$ in Eq.~\eqref{eq:h_eff_2_explicit}. 
The thermal energy from the bath can be used to  induce transitions between the energy levels of the effective Hamiltonian.
Simulations of the dynamics considering the reaction-coordinate Hamiltonian, Eq.~\eqref{eq:h_rc_tot_2}, contain many of these transitions; however, whenever $\Omega \gtrsim \lambda,T$, we expect the effective Hamiltonian to capture the most important transitions. The intuition behind this assumption rests on the fact that the truncation performed on the energy manifold of the reaction coordinate degree of freedom in the polaron frame introduces an error [see Eq.~\eqref{eq:h_eff_total}]. However, we expect this error to remain small whenever the population of the excited states in the reaction-coordinate manifold is small. The physical reasoning behind this assumption is that the reaction-coordinate levels are sufficiently spaced out whenever $\Omega \gtrsim \lambda$ and the population of the highly-excited levels of the reaction coordinate will be small. The second assumption regarding energy scales is that thermal fluctuations will not lead to large populations of the highly-excited states of the reaction coordinate in the polaron frame, that is, $\Omega \gtrsim T$.

Interestingly, it has been demonstrated that the $\Omega \gtrsim \lambda$ condition can be relaxed for certain lattice models such that the effective model leads to an excellent approximation even in the case of arbitrarily large $\lambda$~\cite{Min:2023}. However, for the purpose of evaluating the dynamics in this work, we restrict ourselves to the parameter regime in the energy space where the effective model is straightforwardly justified.

As an example of the power of the effective approach, we show below that we can infer the bath-induced interaction energy between spins from the dynamical signals. While this is difficult in general to achieve with numerical tools (including in the reaction-coordinate treatment), 
the effective Hamiltonian introduces a transparent way to account for inter-spin interactions induced by the bath. 

In light of this, the eigenenergies of the effective Hamiltonian should translate to the most important transitions induced by the bath, which should appear in the transient dynamics. Eq.~\eqref{eq:h_eff_2_explicit} for the $N=2$ case can be diagonalized and we identify the four eigenvalues 
\begin{align}
\label{eq:eigvals_eff}
\omega_{-,\mp} &= \frac{-2\lambda^2 - \sqrt{4\lambda^4 + (\tilde{\Delta}_1 \Omega \mp \tilde{\Delta}_2 \Omega)^2}}{\Omega}, \nonumber \\
\omega_{+,\mp} &= \frac{-2\lambda^2 + \sqrt{4\lambda^4 + (\tilde{\Delta}_1 \Omega \mp \tilde{\Delta}_2 \Omega)^2}}{\Omega},
\end{align}
where $\tilde{\Delta}_{1,2} = e^{-\frac{2\lambda^2}{\Omega^2}} \Delta_{1,2}$. 
Fig.~\ref{fig:1} shows the eigenvalues as a function of $\lambda$ for a specific choice of energy parameters 
in the model. It is to be expected that the relevant transitions and transient dynamics depend on these eigenvalues.

At small $\lambda$, the four eigenvalues can be associated with the energies of the free spins, 
with the two spins pointing down with eigenenergy $\omega_{-,+}$, two single-excitation states with energies $\omega_{-,-}$ and $\omega_{+,-}$,
and the doubly excited state with energy $\omega_{+,+}$, where
\begin{align}
\omega_{-,+}&\xrightarrow{\lambda\to 0} -(\Delta_2+\Delta_1),  \nonumber \\
\omega_{-,-}&\xrightarrow{\lambda\to 0} -\Delta_1+\Delta_2,  \nonumber \\
\omega_{+,-}&\xrightarrow{\lambda\to 0} \Delta_1-\Delta_2,  \nonumber \\
\omega_{+,+}&\xrightarrow{\lambda\to 0} (\Delta_2+\Delta_1).
\end{align}
In this effective Hamiltonian picture, 
if the initial condition only involves the single-excitation manifold, the dynamics would continue to evolve only in this subspace. This is because the interaction with the bath (Eq. \ref{eq:h_eff_total}) will not generate transitions to the other two states for our choice of $\hat{S}$ in Eq.~\eqref{eq:h_and_s_system}.

\section{Results: Thermal equilibrium}
\label{sec:thermaleq}

In this Section, we assess the effective Hamiltonian method in its ability to produce the correct steady-state results for a broad range of system-bath coupling strengths. The effective Hamiltonian method has been benchmarked so far only for equilibrium \cite{Segal:2023Effective,Nick:2023Effective}  and nonequilibrium steady state problems~\cite{Segal:2023Effective}, with a focus on impurity models. 
Lattice models have been examined in Ref.~\cite{Min:2023}.
Here, we focus on system-bath configurations including few spins
and test the capability of the method to capture the polarization behavior throughout the entire range of coupling in the steady-state regime (transients are explored in Sec.~\ref{sec:dynamics_results}).

\begin{figure}[t]
\fontsize{13}{10}\selectfont 
\centering
\includegraphics[width=1\columnwidth]{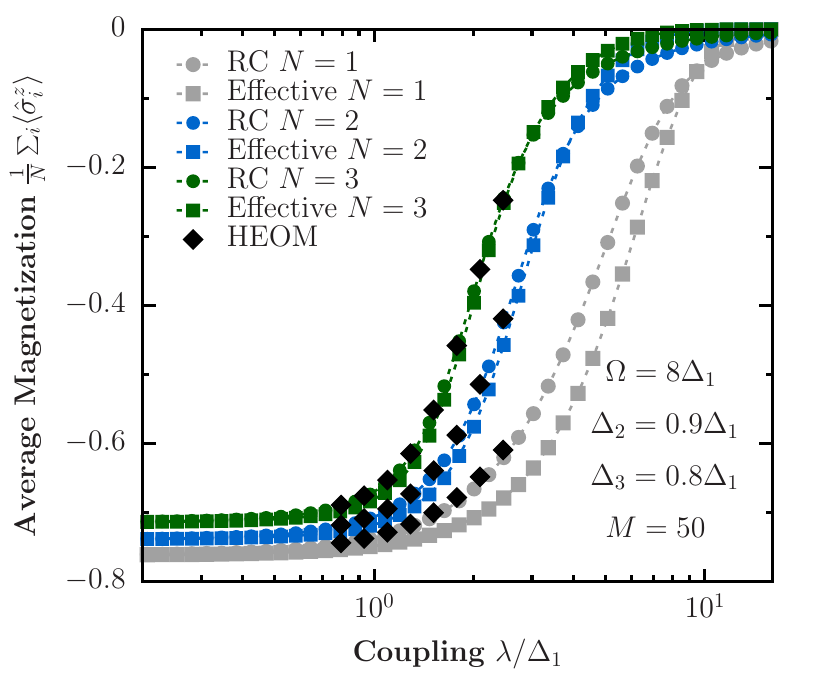}
\caption {Average magnetization at thermal equilibrium for the $N=1$ (grey), $N=2$ (blue) and $N=3$ (green) 
cases as a function of coupling energy $\lambda$. 
Circles denote the calculation for the reaction-coordinate Hamiltonian $\hat{H}_S^{\rm RC}$ ($M = 50$) and squares the corresponding calculation using the effective Hamiltonian $\hat{H}_S^{\rm eff}$. 
The parameters for the calculation are $\Omega = 8\Delta_1$, $\Delta_2 = 0.9\Delta_1$ and $T = \Delta_1$. HEOM calculations are shown with parameters $N_c = 5$ and $N_k = 10$ (see Appendix~\ref{ap:heom}).}
\label{fig:2}
\end{figure}

For a non-perturbative parameter $\lambda$, the equilibrium state will generally deviate from
a thermal Gibbs state~\cite{Nick:2023Effective,Brenes:2023QFI}. 
The equilibrium state, however, will coincide with the state in the limit of long times (steady state) of the dynamics, generated by Eq.~\eqref{eq:redfield_dyn}, 
\begin{align}
\label{eq:lim_infty_state}
\lim_{t \to \infty} \hat{\rho}(t) \to \hat{\rho}_{\rm SS}.
\end{align}
Here $\hat{\rho}(t)$ is the time-dependent density operator of the system. In fact, given that both the reaction-coordinate mapping and the effective Hamiltonian rely on a Markovian embedding, in which the effective or enlarged system Hamiltonian is coupled weakly to a residual reservoir, their equilibrium state is a thermal Gibbs state~\cite{Nick:2023Effective} at the inverse temperature $\beta$ of the bath. In the reaction coordinate picture, the state of the system is
\begin{align}
\label{eq:equilibrium_rc}
\hat{\rho}_{\rm SS}^{\rm RC} =  \frac{       
\Tr_{\rm RC}\left[ e^{-\beta \hat{H}_S^{\rm RC}} \right]}
{Z^{\rm RC}},
\end{align}
where $\hat{H}_S^{\rm RC}$ is the system's RC Hamiltonian from Eq.~\eqref{eq:h_rc_system}, $Z^{\rm RC} = \Tr[e^{-\beta \hat{H}_S^{\rm RC}}]$, and $\Tr_{\rm RC}$ denotes a partial trace over the reaction coordinate (the superscript RC marks the method used). 

We want to probe whether the effective Hamiltonian in Eq.~\eqref{eq:h_eff_2_explicit} provides a good approximation to the polarization at thermal equilibrium, where in this case the state is given by
\begin{align}
\label{eq:equilibrium_eff}
\hat{\rho}_{\rm SS}^{\rm eff} = \frac{e^{-\beta \hat{H}_S^{\rm eff}}}{Z^{\rm eff}},
\end{align}
with $Z^{\rm eff} = \Tr[e^{-\beta \hat{H}_S^{\rm eff}}]$. 

\subsection{Polarization}
\label{sec:polarisation}

The expectation value of the polarization for each of the spins composing the $N=2$ system follows from 
\begin{align}
\langle \hat{\sigma}^z_i \rangle^{\bullet} = \Tr[\hat{\rho}_{\rm SS}^{\bullet} \hat{\sigma}^z_i],
\end{align}
where $\bullet$ denotes either the reaction-coordinate expectation value or its effective Hamiltonian counterpart.
In Fig.~\ref{fig:2}, we present the average magnetization, $\sum_i \langle \hat{\sigma}^z_i \rangle^{\rm RC} / N$ and $\sum_i \langle \hat{\sigma}^z_i \rangle^{\rm eff} / N$ 
at thermal equilibrium as a function of the coupling energy parameter $\lambda$. Together with these results, we complement the calculation with the expectation values obtained from HEOM simulations for certain values of $\lambda$. We make the following observations:

(i) The reaction coordinate calculation, conducted with $M = 50$ levels, yields results consistent with the HEOM calculation, therefore affirming the accuracy of the reaction-coordinate model and method. We note however that the HEOM calculation becomes increasingly more costly for larger values of $\lambda$. 

(ii) Regarding the increase in the number of spins:
In Fig.~\ref{fig:2}, the grey symbols illustrate the polarization results for the $N=1$ case, corresponding to the respective value of $\Delta_1$. The blue symbols represent the results for the $N=2$ system, and the green symbols depict the $N=3$ case. Evidently, as we incorporate more spins, the polarization curve undergoes a shift with respect to $\lambda$ compared to the single spin calculation. This shift implies the presence of bath-induced interactions between spins, which favors ferromagnetic ordering. 

(iii) It appears that the effective Hamiltonian method becomes more accurate with respect to the RC and HEOM approaches moving from $N=1$ to $N=3$. The polarization is already well-approximated, and the behavior is effectively captured in the two-spin model through our effective Hamiltonian treatment.

The second observation above might be taken as an indicator that the interaction energy between spins, $E_I$, could be extracted from this analysis, as the effective Hamiltonian predicts the polarization for the two-spin model at thermal equilibrium. From this, one can compute the equilibrium state using Eq.~\eqref{eq:equilibrium_eff}, as this state replicates the correct polarization as verified by the reaction coordinate state calculation and the HEOM comparison. 
In fact, one can use the eigendecomposition with the eigenvalues in Eq.~\eqref{eq:eigvals_eff} to express the equilibrium state in diagonal form and directly compute the expectation value of the polarization. However, it is clear from Eq.~\eqref{eq:eigvals_eff} that this quantity depends on both the interaction energy $E_I$ and the rescaled spin splittings $\tilde{\Delta}_i$. Indeed, from the equilibrium values one cannot decouple the effect of the environment, as it leads to both interaction between the spins and re-scaled energies of the spins themselves. If one is interested in isolating the energy $E_I$ from the polarization values at equilibrium, then it is required to know how the spin splittings are affected due to environmental effects.

\subsection{Correlations }
\label{sec:correlations}

Before looking at dynamical signals, we here show that the impact of the bath is indeed to generate correlations between each subsystem in the two-spin model at thermal equilibrium. It is also important to understand if the effective Hamiltonian can capture these correlations at thermal equilibrium.

\begin{figure}[t]
\fontsize{13}{10}\selectfont 
\centering
\includegraphics[width=1\columnwidth]{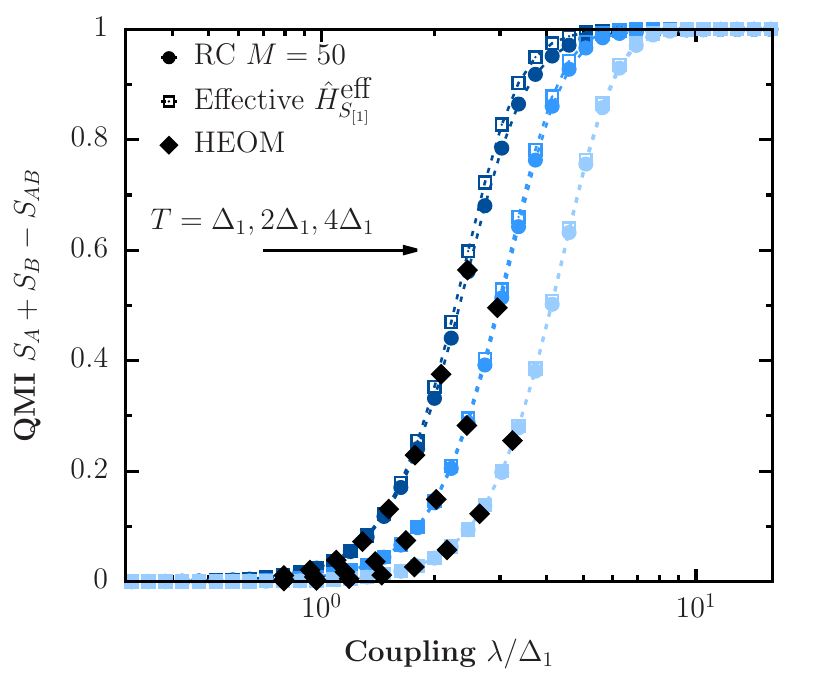}
\caption{Quantum mutual information at thermal equilibrium between each of the spins composing the two-spin system as a function of coupling energy $\lambda$. Circles denote the calculation for the reaction-coordinate Hamiltonian $\hat{H}_S^{\rm RC}$ ($M = 50$) and squares the corresponding calculation using the effective Hamiltonian $\hat{H}_S^{\rm eff}$. 
The parameters for the calculation are $\Omega = 8\Delta_1$ and $\Delta_2 = 0.9\Delta_1$. 
Three values of temperature $T = 1 / \beta$ are shown from left to right: $T = \Delta_1, 2\Delta_1, 4\Delta_1$. HEOM calculations are shown with parameters $N_c = 5$ and $N_k = 10$ (see Appendix~\ref{ap:heom}).}
\label{fig:3}
\end{figure}

To address this objective, we consider again the equilibrium state as described in Eq.~\eqref{eq:lim_infty_state}. We intend to understand if the effective Hamiltonian provides a good approximation of this state and hence the correlations among each subsystem composing the two-spin model. 

We probe the correlations in the equilibrium state using both the reaction-coordinate mapping and the effective Hamiltonian by considering the quantum mutual information (QMI) between each subsystem for the $N=2$ case, i.e., 
\begin{align}
\label{eq:qmi}
I_{(A:B)} = S[\hat{\rho}^A_{\rm SS}] + S[\hat{\rho}^B_{\rm SS}] - S[\hat{\rho}^{AB}_{\rm SS}].
\end{align}
Here, $S[\hat{w}] \defeq -\Tr(\hat{w}\log_2 \hat{w})$ 
is the von Neumann entropy for the state $\hat{w}$ and the indices $A$ and $B$ denote the 
respective state when tracing out the degrees of freedom of the complement subsystem. In our particular case, $A$ and $B$ denote each of the spins composing the system. 
We present results for the QMI as a function of the coupling energy $\lambda$ in Fig.~\ref{fig:3}. 
It can be observed that the effective Hamiltonian provides an accurate description of the correlations at thermal equilibrium compared to the reaction-coordinate Hamiltonian for the parameters chosen. Furthermore, as $\lambda$ increases, correlations between spins $A$ and $B$ increase up to a maximum value. We note that the QMI is a quantity that describes \emph{all} correlations for a given partite state and the correlations observed need not necessarily be quantum in nature, but classical as well. As displayed in Fig.~\ref{fig:3}, the HEOM calculation predicts the same results and trends, therefore validating both the reaction coordinate and the effective model results of the QMI. 

\section{Results: Dynamics}
\label{sec:dynamics_results}

We analyze the out-of-equilibrium dynamics for the cases $N=1$ and  $N=2$ in Sec.~\ref{sec:dynn1} and Sec.~\ref{sec:dynn2}, respectively. For two spins, we investigate their synchronization in 
Sec.~\ref{sec:sync} and the dynamics of correlations in Sec.~\ref{sec:correlations_dyn}. With respect to the dynamics from the HEOM method described in Sec.~\ref{sec:dynamics}, we demonstrate in Appendix~\ref{ap:heom} that  tier (i) and tier (ii) methods, HEOM and RC simulations, respectively, provide the same results. For this reason, for dynamical features we only display results from the cheaper approximation tier (ii). Our focus is in particular on capturing and quantifying bath-induced interactions through the effective Hamiltonian method.

\subsection{Single spin: Relaxation to equilibrium}
\label{sec:dynn1}

\begin{figure}[t]
\fontsize{13}{10}\selectfont 
\centering
\includegraphics[width=1\columnwidth]{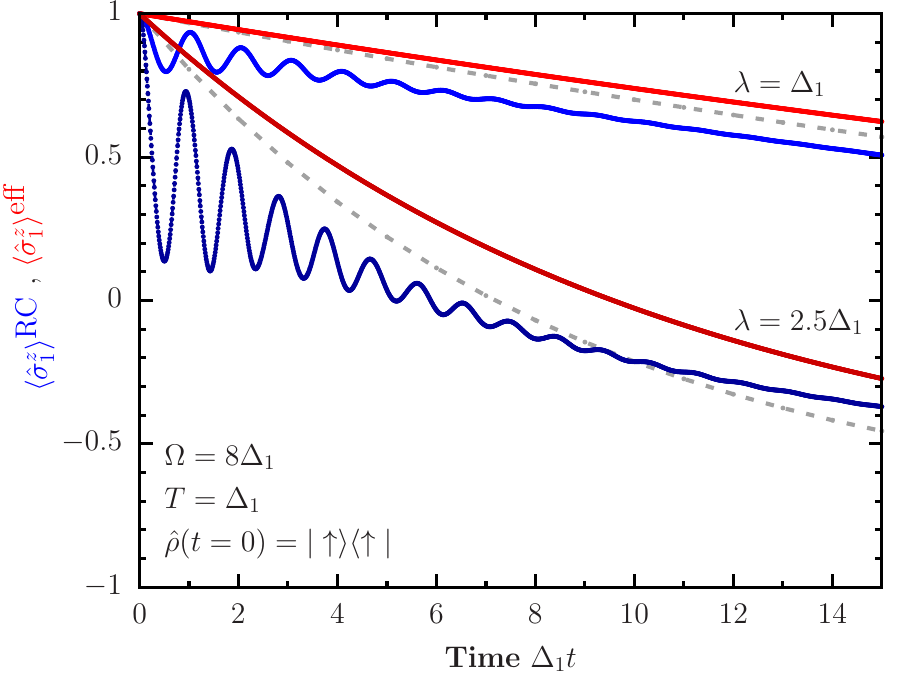}
\caption{Dynamics of the polarization for system ($N=1$) with a coupling parameters $\lambda = \Delta_1$ and $\lambda = 2.5\Delta_1$, as noted in the figure. The blue signals correspond to the reaction coordinate treatment with $M = 25$ while the red lines depict the dynamics using the effective Hamiltonian method. The dynamics are computed from the Redfield QME in Eq.~\eqref{eq:redfield_dyn}. The residual coupling parameter is $\gamma = 0.05$ [see Eq.~\eqref{eq:spec_fun_rc_ohmic}]. The dashed-gray lines depict the dynamics from the weak-coupling approximation.}
\label{fig:4}
\end{figure}

We focus here on the dynamics of a single spin coupled to a bath starting from an out-of-equilibrium state. We recall that Eq.~\eqref{eq:redfield_dyn} dictates the dynamics of the reduced state of a spin system resulting from the interaction with the bath. 

In the strong-coupling regime, the reaction-coordinate mapping allows us to study the dynamics of the spin system with fidelity for the spectral function described in Eq.~\eqref{eq:spec_fun_brownian}, with {\em residual} coupling $\gamma$. To ascertain that the dynamics generated by the reaction-coordinate treatment is correct, we have made a comparison with HEOM calculations in Appendix~\ref{ap:heom}. 
We find that time trajectories coincide. Therefore, we can employ the dynamics obtained from the reaction-coordinate mapping as a point of comparison to analyze the dynamics obtained from the effective Hamiltonian. 
As discussed before, both the reaction-coordinate mapping and the effective Hamiltonian 
treatment allow us to understand strong-coupling effects employing a weak-coupling theory for which the Redfield equation may be justified as a Markovian embedding. 

Our starting point is to choose an initial state in both protocols. For the $N=1$ case, let us consider the decay of a fully-polarized state. Considering that the enlarged system is composed of the spin system plus the RC, a suitable initial state would be a product state between our chosen initial state for the spin system and a thermal Gibbs state for the RC,
\begin{align}
\hat{\rho}^{\rm RC}_S(t = 0) = \ket{\uparrow} \bra{\uparrow} \otimes \frac{e^{-\beta\Omega\hat{a}^{\dagger} \hat{a}}}{\Tr[e^{-\beta\Omega\hat{a}^{\dagger} \hat{a}}]}.
\end{align}
On physical grounds, this is the most sensible choice as we assume that the reaction-coordinate degree of freedom is in thermal equilibrium at the inverse temperature $\beta$ with the residual reservoir. The dynamics of the spin system then follows from the partial trace over the reaction-coordinate degree of freedom. 

From the effective Hamiltonian framework, the initial condition corresponds to 
\begin{align}
\hat{\rho}^{\rm eff}_S(t = 0) = \ket{\uparrow} \bra{\uparrow}.
\end{align}
In both cases, these are the initial states we select to evaluate the dynamics generated by the Redfield  Eq.~\eqref{eq:redfield_dyn} under the reaction-coordinate Hamiltonian $\hat{H}^{\rm RC}_S$ [Eq.~\eqref{eq:h_rc_system}] and the effective Hamiltonian $\hat{H}^{\rm eff}_S$ [Eq.~\eqref{eq:h_eff_explicit}]. 
For the observable, we consider the spin magnetization in the $z$ direction, computed via $\langle \hat{\sigma}^z_i \rangle (t) = \Tr[\hat{\rho}^{\bullet}_S(t) \hat{\sigma}^z_i]$ for the $i$-th spin, where $\hat{\rho}^{\bullet}_S$ is either $\hat{\rho}^{\rm eff}_S(t)$ or $\hat{\rho}^{\rm RC}_S(t)$.

In Fig.~\ref{fig:4} we show the dynamics of the polarization as produced by the reaction coordinate method (blue) compared to the ones obtained from the effective Hamiltonian (red) 
and the weak-coupling Redfield equation limit (gray) for two values of coupling energy, $\lambda = \Delta_1, 2.5\Delta_1$, $T = \Delta_1$ and $\Omega = 8\Delta_1$. 
It is observed that the transient dynamics are {\it not} accurately captured by our effective Hamiltonian model. 
For the case of $N=1$, the observed oscillations in the spin polarization are due to non-Markovian signatures in the relaxation dynamics
that are not included in our effective Hamiltonian. 
We thus find that for the $N=1$ case, while we can understand the equilibrium (long-time) properties from the effective Hamiltonian (Sec.~\ref{sec:polarisation}), the transient dynamics require further refinement of our effective treatment.
Specifically, the complete truncation of the RC manifold down to its ground state erases non-Markovian signatures in the relaxation dynamics. Maintaining more levels in the RC manifold could recover those features.

In Fig.~\ref{fig:4} we also display the weak-coupling dynamics of the $N=1$ Hamiltonian based on Eq.~\eqref{eq:h_and_s_system}. 
Note that, under our mapping, the weak-coupling dynamics is carried out with the Brownian spectral density in Eq.~\eqref{eq:spec_fun_brownian}, as opposed to the Ohmic spectrum in Eq.~\eqref{eq:spec_fun_rc_ohmic}. The weak-coupling dynamics is characterized by pure decay and no oscillations are present.
Notably, when inspecting the relaxation dynamics, the effective Hamiltonian method does not offer 
any advantage over a weak coupling dynamics---though the equilibrium limit is notably distinct.
It was also noted in Ref.~\cite{Nazim} that while at weak-to-intermediate coupling the dynamics of the spin-boson model does not show dramatic signatures, compared to the ultraweak limit, the steady-state values markedly reflect the departure from the weak coupling limit.

\subsection{Two spins: Synchronization and Correlations}
\label{sec:dynn2}

Increasing the number of constituents within the system changes its internal structure and thus the dynamics in a rich way. In particular, from the second term in the right-hand side of Eq.~\eqref{eq:h_eff_2}, it is only the case that $\hat{S}^2 = \mathbf{I}$ for $N=1$. 
However, for general cases where $N > 1$, $\hat{S}^2$ will lead to internal interactions between the spins induced by the bath, which we can understand analytically from the effective Hamiltonian.

\begin{figure}[t]
\fontsize{13}{10}\selectfont 
\centering
\includegraphics[width=0.91\columnwidth]{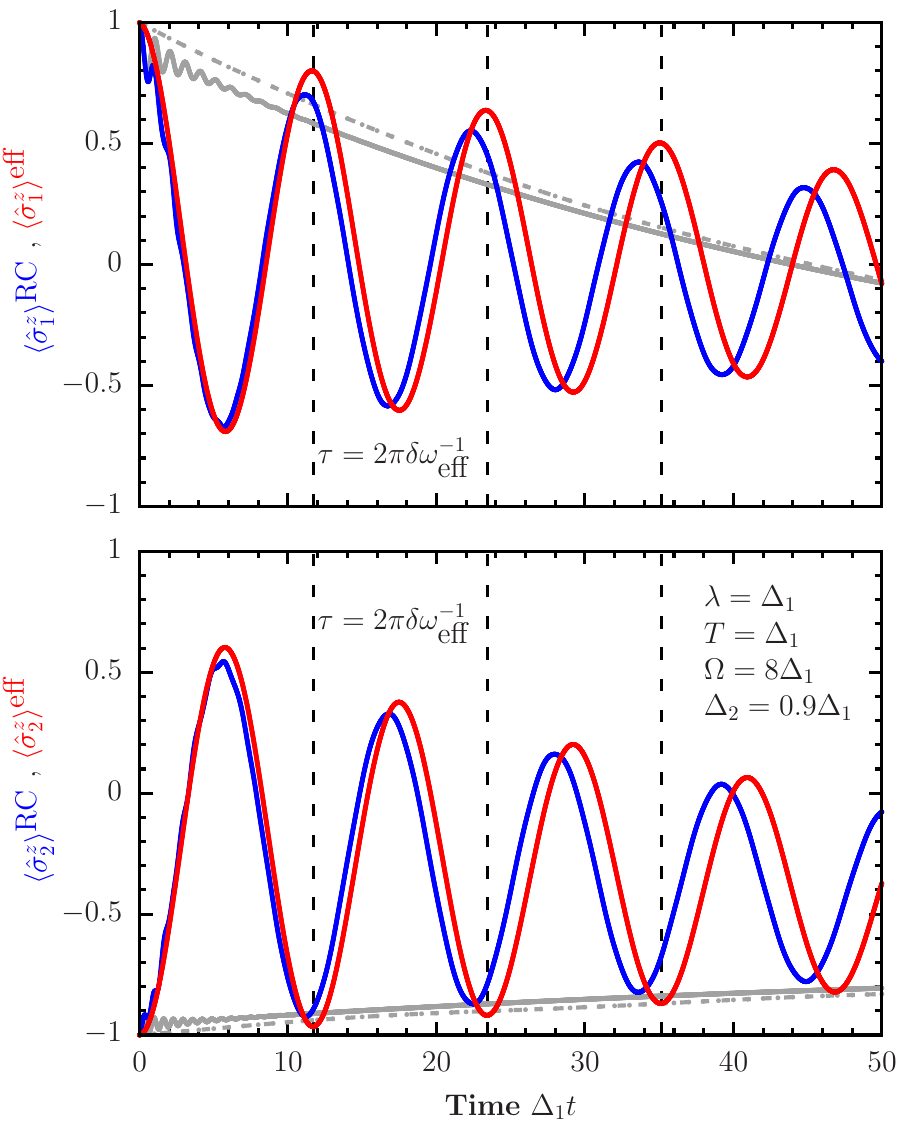}
\caption{Dynamics of the polarization in the two-spin system for the first (top) and second (bottom) spins with a coupling parameter $\lambda = \Delta_1$, demonstrating bath-mediated synchronization of spins. The blue signals correspond to the reaction coordinate treatment with $M = 25$ while the red lines depict the dynamics using the effective Hamiltonian method. The dynamics is computed from the Redfield QME in Eq.~\eqref{eq:redfield_dyn}. The residual coupling parameter is $\gamma = 0.05$ [see Eq.~\eqref{eq:spec_fun_rc_ohmic}]. Vertical lines depict the period obtained from $\tau = 2\pi \delta\omega_{\rm eff}^{-1}$ [see Eq.~\eqref{eq:deltaomega_effective}]. The solid grey curves display the single-spin dynamics as in Fig.~\ref{fig:4}, while the grey-dashed curves display the weak-coupling calculation of the $N=2$ Hamiltonian from Eq.~\eqref{eq:h_and_s_system}.}
\label{fig:5}
\end{figure}

As an example of the utility of the effective Hamiltonian treatment, we examine whether we can extract the bath-mediated interaction energy between spins from the transient dynamics.
While the reaction-coordinate mapping provides no easy way to retrieve this energy scale, below we show that through the effective Hamiltonian mapping we can extract this energy from the effective eigenenergies. 

\begin{figure}[t]
\fontsize{13}{10}\selectfont 
\centering
\includegraphics[width=0.91\columnwidth]{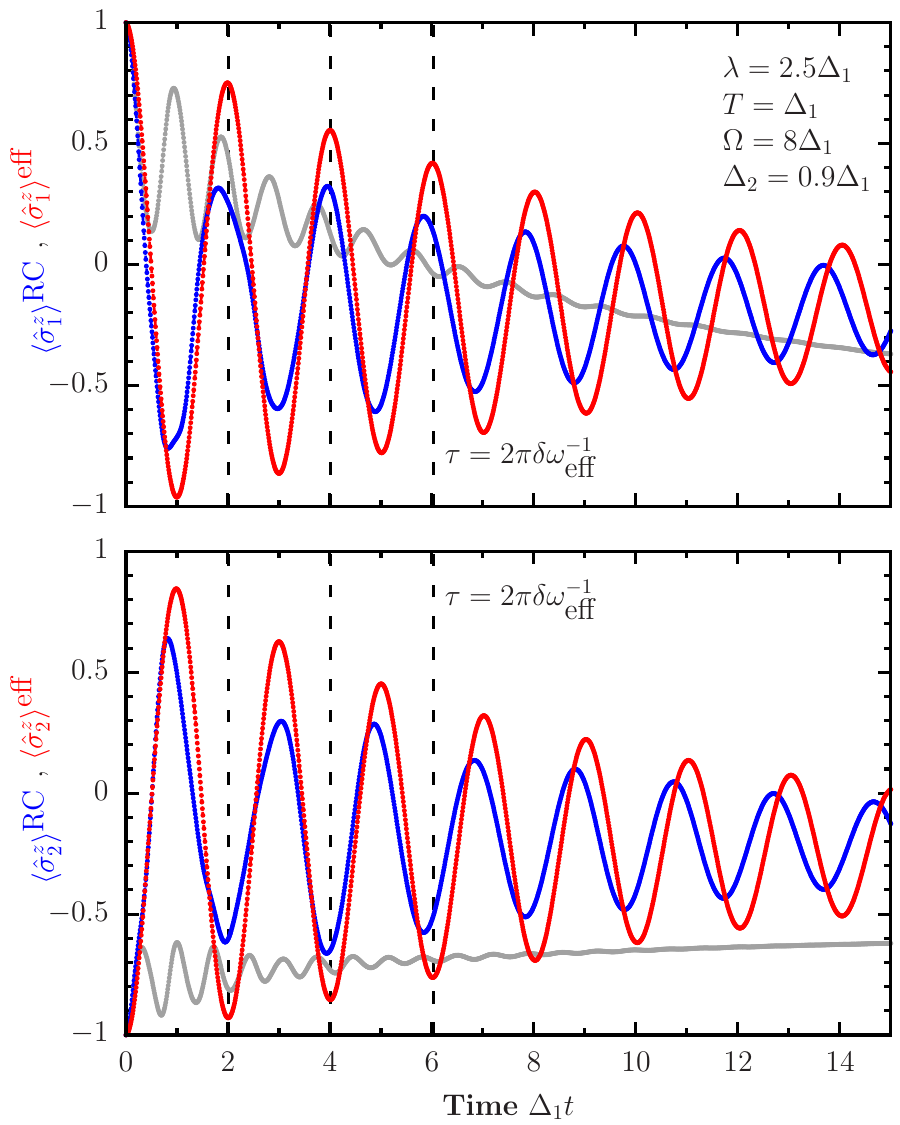}
\caption{Dynamics of the polarization in the two-spin system for the first (top) and second (bottom) spins with a coupling parameter $\lambda = 2.5\Delta_1$, demonstrating synchronization of spins. The blue signals correspond to the reaction coordinate treatment with $M = 25$ while the red lines depict the dynamics using the effective Hamiltonian method. The dynamics is computed from the Redfield QME in Eq.~\eqref{eq:redfield_dyn}. The residual coupling parameter is $\gamma = 0.05$ [see Eq.~\eqref{eq:spec_fun_rc_ohmic}]. Vertical lines depict the period obtained from $\tau = 2\pi \delta\omega_{\rm eff}^{-1}$ [see Eq.~\eqref{eq:deltaomega_effective}]. The grey curves display the single-spin dynamics as in Fig.~\ref{fig:4}.}
\label{fig:6}
\end{figure}

\begin{figure}[t]
\fontsize{13}{10}\selectfont 
\centering
\includegraphics[width=0.91\columnwidth]{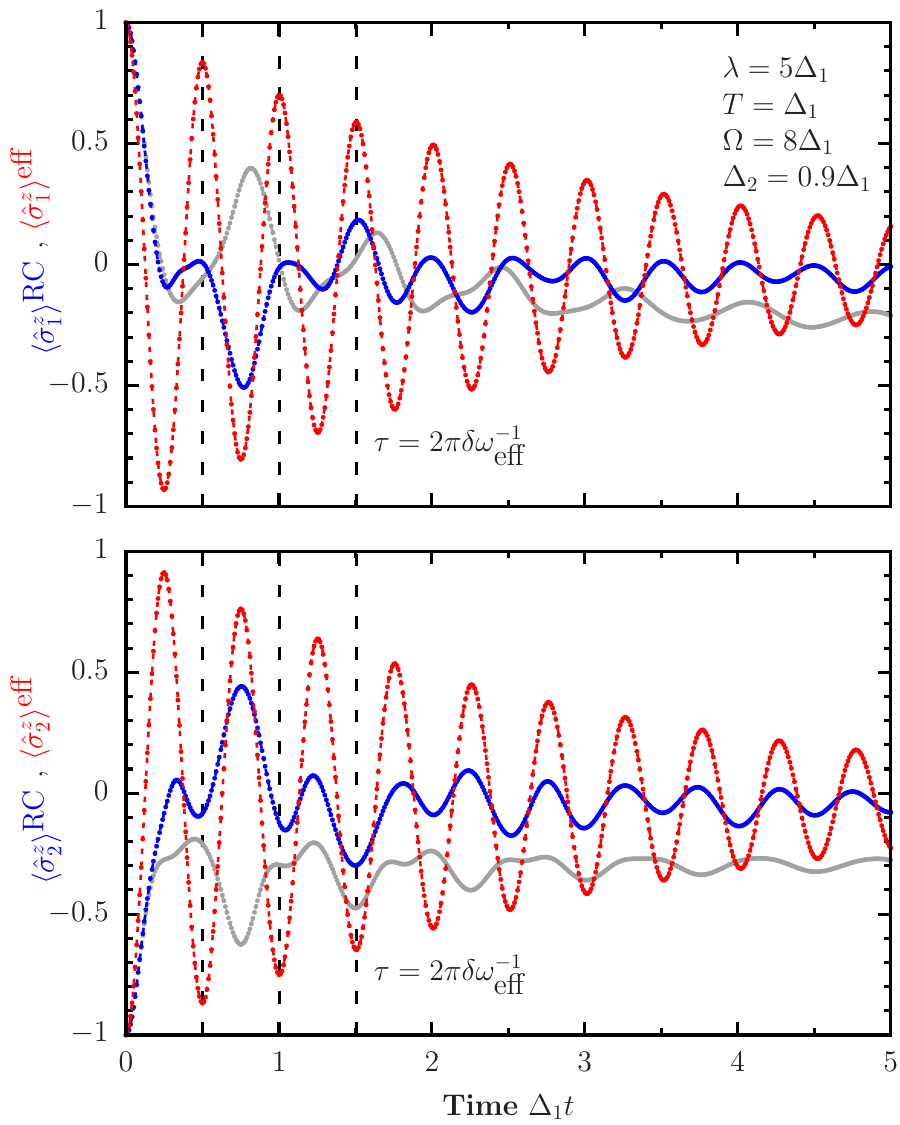}
\caption{Dynamics of the polarization in the two-spin system for the first (top) and second (bottom) spins with a coupling parameter $\lambda = 5\Delta_1$, demonstrating synchronization of spins. The blue signals correspond to the reaction coordinate treatment with $M = 25$ while the red lines depict the dynamics using the effective Hamiltonian method. The dynamics is computed from the Redfield QME in Eq.~\eqref{eq:redfield_dyn}. The residual coupling parameter is $\gamma = 0.05$ [see Eq.~\eqref{eq:spec_fun_rc_ohmic}]. Vertical lines depict the period obtained from $\tau = 2\pi \delta\omega_{\rm eff}^{-1}$ [see Eq.~\eqref{eq:deltaomega_effective}]. The grey curves display the single-spin dynamics as in Fig.~\ref{fig:4}.}
\label{fig:7}
\end{figure}

We start by choosing an initial state for the $N=2$ case. For reasons that shall become apparent later, we consider an initial state that is fully polarized 
in the up direction of the $z$ component for the first spin and down for the second, 
\begin{align}
\label{eq:init_state_eff_2}
\hat{\rho}^{\rm eff}_S(t = 0) = \ket{\uparrow \downarrow} \bra{\uparrow \downarrow}.
\end{align}
This is the initial state we choose to evaluate the dynamics generated by the Redfield equation under the effective Hamiltonian $\hat{H}^{\rm eff}_S$ from Eq.~\eqref{eq:h_eff_2_explicit}. 
Our point of comparison is the dynamics generated by the reaction-coordinate Hamiltonian $\hat{H}^{\rm RC}_S$ from Eq.~\eqref{eq:h_rc_system}, for which we need to select a suitable initial state. As before, the relevant state in this case is
\begin{align}
\label{eq:init_state_rc_2}
\hat{\rho}^{\rm RC}_S(t = 0) = \ket{\uparrow \downarrow} \bra{\uparrow \downarrow} \otimes \frac{e^{-\beta\Omega\hat{a}^{\dagger} \hat{a}}}{\Tr[e^{-\beta\Omega\hat{a}^{\dagger} \hat{a}}]}.
\end{align}
Our objective is to evaluate the dynamics generated by the effective Hamiltonian and compare them with exact results. In Appendix~\ref{ap:heom}, we show that the dynamics generated from the full reaction-coordinate Hamiltonian coincide with those generated with HEOM (tiers (i) and (ii) from Sec.~\ref{sec:dynamics}). For this reason, we shall consider the full reaction-coordinate Hamiltonian as exact results.

\subsubsection{Synchronization behavior and its temperature dependence}
\label{sec:sync}

In Fig.~\ref{fig:5}, Fig.~\ref{fig:6} and Fig.~\ref{fig:7}, we display the dynamics 
of $\langle \hat{\sigma}^z_i \rangle (t)$ for the values of $\lambda = \Delta_1$, $\lambda = 2.5\Delta_1$ and $\lambda = 5\Delta_1$; respectively. 
For these calculations, we chose $\Omega = 8\Delta_1$ and $\Delta_2 = 0.9\Delta_1$, for a particular value of temperature $T = \Delta_1$. The blue curves depict the dynamics generated by the reaction-coordinate Hamiltonian, i.e., the exact results; while the red curves display the dynamics generated with the effective Hamiltonian model.

The dynamics is characterized by oscillations until the stationary value for the magnetization is reached. We also display the dynamics of the $N=1$ case with the specific value of $\Delta_i$ for the $i$-th spin. Trivially, synchronization does not occur in this case due to the absence of an interaction with any other spin degree of freedom
This can be observed in the solid-gray curves in Fig.~\ref{fig:5}, Fig.~\ref{fig:6} and Fig.~\ref{fig:7}. Furthermore, in Fig.~\ref{fig:5}, we display the dynamics using the weak-coupling approach for the $N=2$ model with dashed-gray curves, which neither posses the aforementioned oscillations in the spin polarization due to spin-bath effects, nor the synchronization effect observed due to strong-coupling and bath-mediated interactions \cite{Lamb}.
As a reminder, the weak-coupling calculation is carried out with the Brownian spectral density function in Eq.~\eqref{eq:spec_fun_brownian}, as opposed to the Ohmic spectrum achieved after the RC mapping, Eq.~\eqref{eq:spec_fun_rc_ohmic}.

We find that the transient dynamics is well-approximated by our effective Hamiltonian treatment in the regime of weak to intermediate coupling. Most importantly, Fig.~\ref{fig:5} and Fig.~\ref{fig:6} demonstrate
that the effective Hamiltonian treatment captures both the frequency and magnitude of the bath-induced synchronization dynamics. At stronger coupling, Fig.~\ref{fig:7} shows that while the effective method properly captures the period of synchronization, it overestimates the magnitude of oscillations, or in other words, it underestimates the dissipative dynamics. Synchronization, in this case, becomes apparent from the common frequency of the magnetization oscillations in time for both spins. Such synchronized dynamics arise from the interaction induced by the bath and not due to internal degrees of freedom.

The period $\tau$ of oscillations is connected to the relevant transition of the effective eigenenergies, given by
\begin{align}
\tau = 2\pi \delta \omega^{-1}_{\rm eff},
\end{align}
where $\delta \omega^{-1}_{\rm eff}$ can be computed analytically from Eq.~\eqref{eq:eigvals_eff},
\begin{align}
\label{eq:deltaomega_effective}
\delta \omega_{\rm eff} = \omega_{+,+} - \omega_{-,+} = 2\sqrt{\left(\frac{2\lambda^2}{\Omega}\right)^2 + \left( \tilde{\Delta}_1 - \tilde{\Delta}_2 \right)^2}.
\end{align}
In the case of spins of similar splittings, $\Delta_1 \approx \Delta_2$, as we have considered in simulations, the second term in Eq.~(\ref{eq:deltaomega_effective}) can be neglected compared to the first. We then directly connect the interaction energy between the spins to the period of oscillation 
observed in Figs.~\ref{fig:5}-\ref{fig:7},
\begin{align}
\label{eq:e_i_lambda}
\delta \omega_{\rm eff} \approx \frac{4\lambda^2}{\Omega} = 2E_I.
\end{align}
In arriving at Eq.~\eqref{eq:deltaomega_effective}, 
we recall that, as $\lambda \to 0$, each of the four eigenvalues of the effective Hamiltonian correspond to the eigenstates $\ket{\downarrow \downarrow}$, $\ket{\uparrow \downarrow}$, $\ket{\downarrow \uparrow}$ 
and $\ket{\uparrow \uparrow}$ (see Eq.~\eqref{eq:eigvals_eff} and Fig.~\ref{fig:1}). 
These eigenvalues change as $\lambda$ increases due to the effect of the bath. Following $\hat{H}^{\rm{eff}}_S$ from Eq.~\eqref{eq:h_eff_2_explicit} and $\hat{S} = \hat{\sigma}^x_1 + \hat{\sigma}^x_2$, after ignoring constant terms to the Hamiltonian we have
\begin{align}
\hat{H}^{\rm{eff}}_S &= e^{-\frac{2\lambda^2}{\Omega^2}} \left( \Delta_1 \hat{\sigma}^z_1 + \Delta_2 \hat{\sigma}^z_2 \right) - \frac{2\lambda^2}{\Omega} \hat{\sigma}^x_1 \hat{\sigma}^x_2,
\end{align}
where the second term corresponds to the interaction induced by the common reservoir. This Hamiltonian is $\mathbb{Z}_2$ symmetric, such that the operator $\hat{O} = \prod_{i=1}^{N} \hat{\sigma}^z_i$ commutes with the Hamiltonian. This implies that the bath effective interaction energy can induce spin-flipping interactions on both spins but it does not admix spin inversion sectors. In light of this, the choice of initial condition becomes important. With our preparation, $\hat{\rho}(t = 0) = \ket{\uparrow \downarrow}\bra{\uparrow \downarrow}$, the relevant mode presented in the dynamics will be the one given by the difference between the states corresponding to $\ket{\uparrow \downarrow}$ and $\ket{\downarrow \uparrow}$, both of which belong to the same spin inversion subsector. In such a way that the frequency of oscillations must be related to $\delta \omega_{\rm eff}$ in Eq.~\eqref{eq:deltaomega_effective}. 

\begin{figure}[t]
\fontsize{13}{10}\selectfont 
\centering
\includegraphics[width=0.91\columnwidth]{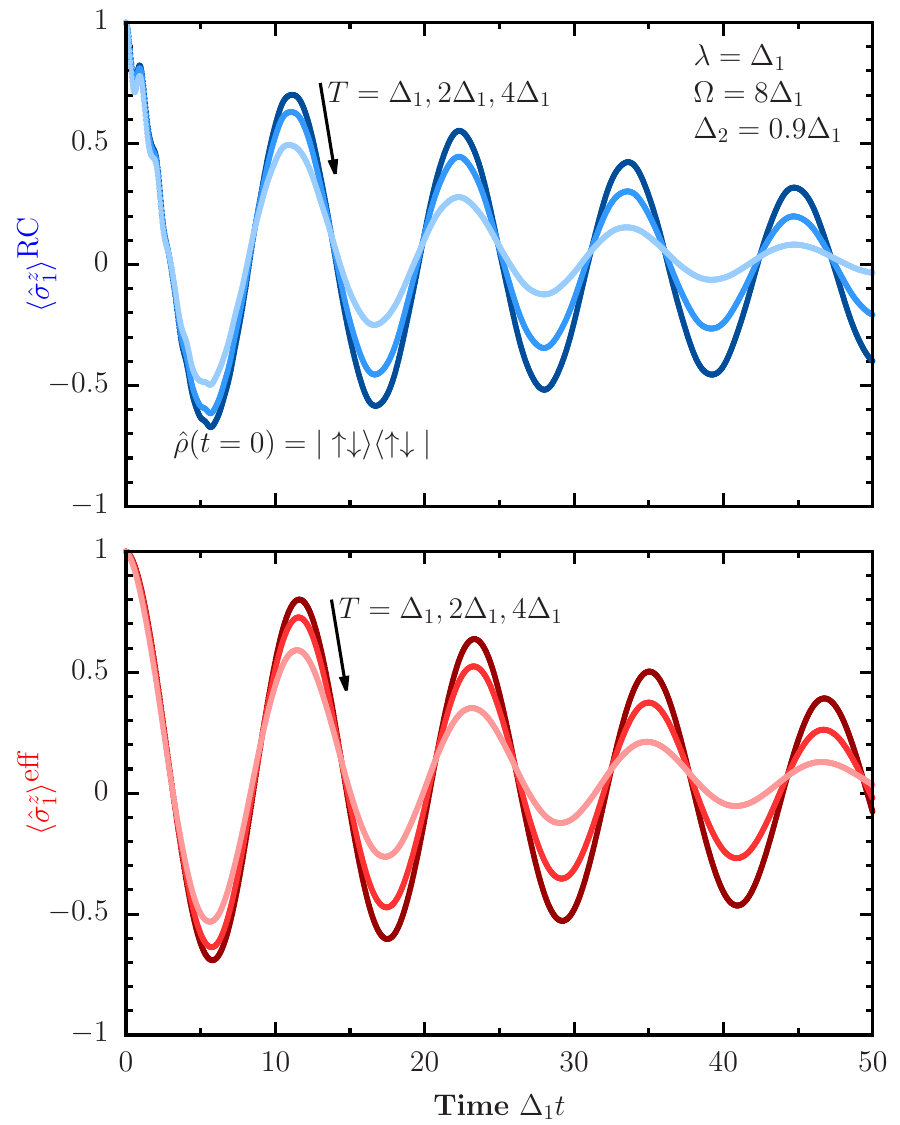}
\caption{Dynamics of the polarization in the two-spin system for the first spin with spin-splitting $\Delta_1$ obtained from the (top) reaction-coordinate mapping ($M = 25$) and the (bottom) effective Hamiltonian treatment for three different values of temperature $T = \Delta_1, 2\Delta_1, 4\Delta_1$. Calculations are shown for $\lambda = \Delta_1$. The residual coupling parameter is $\gamma=0.05$, [see Eq.~\eqref{eq:spec_fun_rc_ohmic}].}
\label{fig:8}
\end{figure}

We have found that the effective Hamiltonian model, while capable of approximating the multi-spin synchronized dynamics at weak-to-moderate couplings (Fig.~\ref{fig:5}-\ref{fig:6}) for special (experimentally-relevant) initial conditions, fails to replicate the exact dynamics at stronger coupling. 
However, while the amplitude of oscillations is not appropriately captured by the effective Hamiltonian, the periodicity in oscillations is correct, and it gives an analytical understanding of relevant energy scales. 

From Eq.~\eqref{eq:e_i_lambda}, we observe that these oscillations are characterized by a frequency which is {\it temperature independent}.

To test this prediction, in Fig.~\ref{fig:8} we study the polarization of the first spin for the $N=2$ model, with the same initial condition we considered above and the same energetic parameters. Each of the curves displayed shows the dynamics with different temperatures, $T = \Delta_1, 2\Delta_1, 4\Delta_1$. In Fig.~\ref{fig:8}(top), we show the true dynamics from the reaction-coordinate Hamiltonian $\hat{H}^{\rm RC}_S$ Eq.~\eqref{eq:h_rc_system}, and in Fig.~\ref{fig:8}(bottom) we present analogous results using the effective Hamiltonian $\hat{H}^{\rm eff}_S$ Eq.~\eqref{eq:h_eff_2_explicit}. While the approach towards equilibration is faster for higher temperatures, the frequency of the characteristic oscillations does not change with temperature, and it can be inferred analytically from the effective Hamiltonian, Eq.~\eqref{eq:e_i_lambda}. 
Altogether, the effective Hamiltonian theory correctly captures the effect of bath-induced spin synchronization and its temperature dependence, extending previous studies that were limited to zero temperature~\cite{Orth:2010}.

\subsubsection{Dynamics of correlations}
\label{sec:correlations_dyn}

In Sec.~\ref{sec:correlations}, we have observed that the effect of the bath onto the two-spin system $N=2$ is to generate correlations at thermal equilibrium. It is of interest to consider the dynamics of these correlations and, in particular, to analyze the nature of these correlations.

From the QMI in Eq.~\eqref{eq:qmi}, we can gather the behavior of correlations. However, in the QMI all correlations are taken into account and one cannot classify whether these correlations are classical or non-classical. For the purposes of the classification of correlations, we now introduce in our calculations the dynamics of the entanglement negativity~\cite{Karol:1998Neg,Vidal:2002Neg}, defined as
\begin{align}
\label{eq:neg_1}
N^{\bullet}_{(A:B)}(t) = \frac{|| \hat{\rho}^{\bullet}_{T_A}(t) ||_1 - 1}{2},
\end{align}
where the $\bullet$ denotes the density operator on which we compute the entanglement negativity, i.e., the RC or effective Hamiltonian from tier (ii) and (iii) in Sec.~\ref{sec:dynamics}. $|| \hat{w}_{T_A}(t) ||_1$ corresponds to the trace norm of the operator $\hat{w}$ as a function of time under the partial transpose operation under subsystem $A$ (i.e., the first spin of the two-spin system). 
Consequently, the entanglement negativity may also be computed from the sum of the negative eigenvalues $\kappa^{\bullet}_i(t)$ of $\hat{\rho}^{\bullet}_{T_A}(t)$ with the relation~\cite{Khandelwal:2020Neg}
\begin{align}
\label{eq:neg_2}
N^{\bullet}_{(A:B)}(t) = \sum_{\kappa^{\bullet}_i(t) < 0} |\kappa^{\bullet}_i(t)|.
\end{align}
The intuition behind the entanglement negativity as an entanglement measure is related Peres-Horodecki~\cite{Peres:1996Neg,Horodecki:1996Neg} criterion for separability. Fundamentally, the degree to which $\hat{\rho}^{\bullet}_{T_A}(t)$ fails to be positive provides a quantitative indication of separability. In this respect, we can classify correlations between classical and non-classical with this measure as opposed to the QMI, in which these correlations are intertwined. 

\begin{figure}[t]
\fontsize{13}{10}\selectfont 
\centering
\includegraphics[width=0.91\columnwidth]{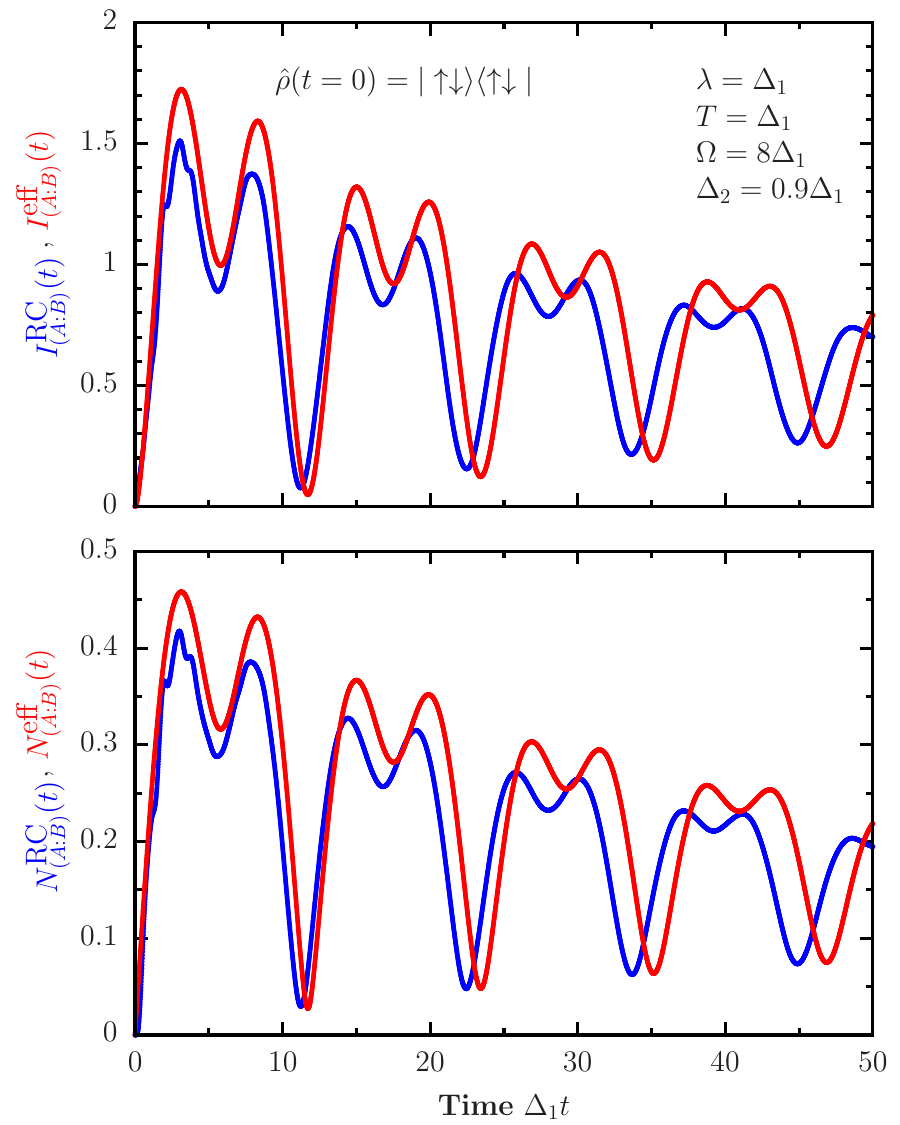}
\caption{Dynamics of the QMI (top) and entanglement negativity (bottom) as a function of time for the two-spin system ($\Delta_2 = 0.9\Delta_1$) obtained from the (blue) reaction-coordinate mapping ($M = 25$) and the (red) effective Hamiltonian treatment for $T = \Delta_1$ and $\lambda = \Delta_1$. Calculations are shown for $\lambda = \Delta_1$. The dynamics is computed from the Redfield QME in Eq.~\eqref{eq:redfield_dyn}. The residual coupling parameter is $\gamma=0.05$, [see Eq.~\eqref{eq:spec_fun_rc_ohmic}].}
\label{fig:9}
\end{figure}

In Fig.~\ref{fig:9} we display the dynamics of both the QMI and the entanglement negativity as a function of time for $\lambda = \Delta_1$ and $T = \Delta_1$ for the two-spin case $N=2$ using the same initial states as considered before [Eq.~\eqref{eq:init_state_eff_2} and Eq~\eqref{eq:init_state_rc_2} for the effective and RC cases, respectively]. In both cases we examine the correlations between spin 1 and spin 2, denoted here by  
$A$ and spin $B$, respectively.
We first note that both the QMI and $N^{\bullet}_{(A:B)}(t)$ displays oscillating features as the polarization dynamics (Fig.~\ref{fig:6},~\ref{fig:7},~\ref{fig:8}), however, the period of the oscillations is not the same. With respect to the accuracy of the effective model, as for the dynamics of the polarization, we can approximate the dynamics at weak-to-intermediate coupling energy with this theory.

Most importantly, we  see that the dynamics reveal that there are correlations developing in the transient states from the QMI [Fig.~\ref{fig:9}(top)], which prevail even in the limit of long times (see Fig.~\ref{fig:3}). Interestingly, from the entanglement negativity [Fig.~\ref{fig:9}(bottom)], it is observed that some of these correlations as a function of time are non-classical in nature, meaning that they stem from the non-separability of the reduced state between spin $A$ and spin $B$. However, from the computation of the entanglement negativity of the {\it equilibrium states} (an analogous calculation of the entanglement negativity as in Fig.~\ref{fig:3}) we have obtained that in the equilibrium states the entanglement negativity is zero. This immediately implies that all the correlations in Fig.~\ref{fig:3} are classical correlations. There are non-classical correlations in the transient states,  as observed in Fig.~\ref{fig:9}(bottom), but they are washed away in the limit of long times. The presence of these non-classical correlations may be relevant to the engineering of entangled states for quantum computation. 
We highlight that these intriguing observations may depend on temperature, a question that we leave for future studies. 

On a final note, we point out that the dynamics of the entanglement negativity displayed in Fig.~\ref{fig:9} from the RC method using the Redfield Eq.~\eqref{eq:redfield_dyn} coincide with the dynamics computed from the HEOM, validating the RC results displayed (see Appendix~\ref{ap:heom}).


\section{Assessment of the effective Hamiltonian method}
\label{sec:assess}

We summarize the regime of applicability of the effective Hamiltonian theory. The effective Hamiltonian method prepares a system Hamiltonian that incorporates strong coupling effects through  renormalization of energy parameters and the generation of bath-induced interaction terms. 
As such, by inspecting the effective Hamiltonian one can gain a direct understanding of expected strong coupling signatures. On the other hand, the coupling of the effective Hamiltonian to the residual bath is assumed weak and it is handled at the level of the Born-Markov approximation. Therefore, non-Markovian relaxation effects in the effective Hamiltonian dynamics are not taken into account. Based on this understanding, we summarize our assessment of the effective Hamiltonian theory:

(i) The theory properly captures the {\it steady state} behavior of spins in a boson bath, from the ultraweak to the ultrastrong coupling limits. We speculate that the method becomes more accurate when increasing the number of spins in the system at the level of equilibrium state description. It requires further assessment to understand the validity of the effective Hamiltonian method as the thermodynamic limit $N\to\infty$ is approached.

(ii) In the transient regime, the effective Hamiltonian method correctly captures the effect of bath-induced interaction between spins. This includes phenomenon such as the synchronization of spins, where the method accurately reproduces the frequency and magnitude of this effect, particularly at weak to intermediate couplings. However, as mentioned in point (i), the effective Hamiltonian treatment, with its truncation of the RC, has limitations. It is unable to capture secondary details in the relaxation dynamics that correspond to non-Markovian effects with respect to the residual bath. 

Given its underlying approximations, the effective Hamiltonian method is useful when the temperature of the bath is comparable to energy levels in the system, while the characteristic frequency of the bath is high, i.e., $\Omega \gtrsim \Delta,T$. However, as was demonstrated in Ref.~\cite{Nick:2023Effective}, performing the polaron mapping in Eq.~\eqref{eq:polaron} in a variational manner can extend the reach of the method to handle cases with $\Omega$ comparable to $\Delta$ and $T$.

\section{Conclusions}
\label{sec:conclusion}

The transient dynamics of initially out-of-equilibrium system-bath configurations at strong coupling is an intricate problem, which is often addressed through complex and expensive numerical techniques. The difficulty arises from the competing energy scales between coherent and incoherent effects induced by internal Hamiltonian dynamics and environmental effects. In this work, we have taken a step forward towards understanding such complex dynamics using a semi-analytical treatment. We have understood the problem from the perspective of the effective Hamiltonian theory. This framework embeds the action of the bath onto internal degrees of freedom via a mapped effective Hamiltonian. 

In delimiting the scope of the problem, we have adopted specifically tailored initial states and a bath spectral density function. Our working assumption is the one required by the applicability of a Markovian embedding via the reaction-coordinate mapping~\cite{Iles:2014RC}. This assumption requires specific spectral density functions such that after the mapping the residual reservoir is coupled weakly to the newly-mapped Hamiltonian. Aside from that, the assumption on the energy scale imposed by the bath characteristic energy $\Omega \gtrsim \lambda, T$ remains valid in our calculations. Interestingly, such condition may be relaxed towards understanding equilibrium states as was recently demonstrated in Refs.~\cite{Segal:2023Effective,Nick:2023Effective,Min:2023}.

We have first observed that starting from an experimentally-relevant out-of-equilibrium initial product state between the system and the bath; and due to the coherent effects induced by the bath, the spin polarization is characterized by oscillating signals. We have found that for a single spin ($N=1$) these oscillations are not captured by the theory of effective Hamiltonian and only equilibrium properties are well-approximated. This is due to the fact that for a single spin degree of freedom, such oscillations stem from a backflow from the bath, which we do not capture with an effective Hamiltonian. However, considering a larger system ($N=2$), the bath introduces couplings between each subsystem, which is appropriately captured by our analytical treatment. At weak to intermediate coupling energies, the effective Hamiltonian appropriately approximates such signals. We recovered the frequency of the oscillations, which we found to be directly proportional to the interaction energy between each spin composing the system. 

We have found that the coupling of a two-spin system to a common thermal reservoir leads to correlations between each spin even if the spins are not directly coupled to each other. Interestingly, at equilibrium these are all classical correlations but the transient states develop non-classical correlations stemming from non-separability. As a function of temperature, non-classical correlations between each spin subsystem may develop at low temperature in the steady state regime. We leave an investigation of the temperature dependence of correlations to future work.

Our results may be extended by considering a less extreme truncation scheme for the reaction coordinate manifold.  Instead of retaining only the ground state of the reaction coordinate, one may keep two or more levels in the effective Hamiltonian. This extended treatment is expected to improve on the predicted dynamics. However, it will complicate the procedure and its analysis, requiring working with a larger space and tracing out the reaction coordinate as part of the procedure. More broadly, it is interesting to explore alternative mapping approaches complementary to the effective Hamiltonian: methods that are based on the Schriffer-Wolff transformation, chain mapping, or alternative decomposition of bath modes, see e.g. Refs. \cite{Prior:2010,Chin:2010,Woods:2014Embedding,Ashida1,Ashida2,Nori23}.

The results presented here have a strong connection to various experimental platforms, for which similar scenarios may occur and may have significant implications. 
These include superconducting circuits, for which qubits may be coupled through multimode waveguides \cite{superQbits19}, as well as hybrid nanomechanical or optomechanical systems, which naturally connect harmonic modes with two-level systems (e.g., localized defects, atoms) \cite{Shabir}. We are specifically interested in solid-state defect systems (such as nitrogen-vacancy centers in diamond), and the coupling of such spin defects through a common bus (e.g., phononic modes in a structured crystal such as surface acoustic waves) \cite{BarGill12}. 
Effective coherent coupling between these defects is a long-standing open problem in the community, as it is a major obstacle on the path toward realizing useful quantum processing and quantum simulation in these platforms. An interesting extension of the current work could consider other types of environments mediating the coupling between the spins, namely spin baths. Addressing this question is interesting both on a fundamental level and in terms of practical implications on relevant quantum technologies.

\begin{acknowledgments}
The work of M.B.~has been supported by the Centre for Quantum Information and Quantum Control (CQIQC) at the University of Toronto and an NSERC Discovery grant of D.S. The work of B.M. was partially supported by an Alliance International Catalyst Quantum grant. The work of N.A.-S. was supported by an Ontario Graduate Scholarship.
D.S.~acknowledges support from NSERC and the Canada Research Chair program. 
N.B. acknowledges support from the European Union’s Horizon 2020 research and innovation program under Grant Agreements No. 101070546 (MUQUABIS) and No. 828946 (PATHOS), and has been supported in part by the Ministry of Science and Technology, Israel, the innovation authority (Project No. 70033), and the ISF (Grants No. 1380/21 and No. 3597/21).
Computations were performed on the Niagara supercomputer at the SciNet HPC Consortium. SciNet is funded by: the Canada Foundation for Innovation; the Government of Ontario; the Ontario Research Fund - Research Excellence; and the University of Toronto. 
\end{acknowledgments}

\appendix

\section{Derivation of the Effective Hamiltonian for the two-spin model}
\label{ap:eff_h}

Following Eq.~\eqref{eq:h_eff_2}, for the specific case of $\hat{H}_S$ and $\hat{S}$ in Eq.~\eqref{eq:h_and_s_system}, we have
\begin{align}
\hat{\tilde{H}}_S &= e^{\frac{\lambda}{\Omega}(\hat{a}^{\dagger} - \hat{a})\hat{S}} \hat{H}_S e^{-\frac{\lambda}{\Omega}(\hat{a}^{\dagger} - \hat{a})\hat{S}} \nonumber \\
&= \sum_{i=1}^{N} \Delta_i e^{\hat{A}\hat{\sigma}^x_i} \hat{\sigma}^z_i e^{-\hat{A}\hat{\sigma}^x_i},
\end{align}
where $\hat{A} = \frac{\lambda}{\Omega}(\hat{a}^{\dagger} - \hat{a})$. Each of these terms may be computed through a Baker–Campbell–Hausdorff (BCH) expansion~\cite{BreuerPetruccione} to obtain
\begin{align}
\label{eq:polaron_zz}
e^{\hat{A}\hat{\sigma}^x_i} & \hat{\sigma}^z_{i} e^{-\hat{A}\hat{\sigma}^x_i} \nonumber \\
&= \hat{\sigma}^z_i + \hat{A} \left[ \hat{\sigma}^x_i, \hat{\sigma}^z_i \right] + \frac{1}{2!} \hat{A}^2 \left[ \hat{\sigma}^x_i, \left[ \hat{\sigma}^x_i, \hat{\sigma}^z_i \right] \right] + \cdots \nonumber \\
&= \frac{e^{2\hat{A}}(\hat{\sigma}^z_i - {\rm i} \hat{\sigma}^y_i) + e^{-2\hat{A}}(\hat{\sigma}^z_i + {\rm i} \hat{\sigma}^y_i)}{2}.
\end{align}
The next step in the procedure to generate the effective system Hamiltonian is to truncate $\hat{\tilde{H}}_S$ to the ground-state manifold of the reaction coordinate. This may be better understood from the properties of the displacement operator usually found in the context of coherent states, i.e., $\hat{D}(\alpha) = e^{\alpha (\hat{a}^{\dagger} - \hat{a})}$, which satisfies $\hat{D}^{\dagger}(\alpha) = \hat{D}(-\alpha)$. A coherent state can be generated by the displacement operator applied to the vacuum state $\ket{\alpha} = \hat{D}(\alpha) \ket{0}$, where
\begin{align}
\ket{\alpha} = e^{-|\alpha|^2 / 2} \sum_{k=1}^{\infty} \frac{\alpha^k}{\sqrt{k!}} \ket{k}.
\end{align}
It follows that $\braket{0 | e^{\hat{A}} | 0} = \braket{0 | e^{-\hat{A}} | 0} = \braket{0 | \alpha} = e^{-\lambda^2 / 2\Omega^2}$. With these results at hand, we find
\begin{align}
\label{eq:h_eff_heisenberg_projected}
\hat{H}^{\rm eff}_S &= \hat{Q}_0 \hat{U} \hat{H}_S \hat{U}^{\dagger} \hat{Q}_0 - \frac{\lambda^2}{\Omega}\hat{S}^2 \nonumber \\
&= e^{-\frac{2\lambda^2}{\Omega^2}} \hat{H}_S - \frac{\lambda^2}{\Omega} \hat{S}^2.
\end{align}

\section{Bath-induced interactions from a polaron transformation}
\label{ap:polaron}

In this Appendix, we provide an alternative approach to obtaining the effective Hamiltonian via the polaron transformation for the $N=2$ case. Starting with the original Hamiltonian of two spins immersed in a common bath, without a direct interaction term,
\begin{equation}
\begin{aligned}
    \hat{H} = \Delta_1\hat{\sigma}^z_1+\Delta_2\hat{\sigma}^z_2+\sum^2_{i=1}\hat{\sigma}^x_i\sum_kt_{k,i}(\hat{c}^\dagger_k+\hat{c}_k)+\sum_k\nu_k\hat{c}^\dagger_k\hat{c}_k,
    \end{aligned}
\end{equation}
we perform consecutive polaron transformations with $U_1$ and $U_2$, where $U_i=\exp(-{\rm{i}}\hat{\sigma}^x_i\hat{B}_i/2)$ and $\hat{B_i}=2{\rm{i}}\sum_k\frac{f_{k,i}}{\nu_k}(\hat{c}^\dagger_k-\hat{c}_k)$. 
These transformations are referred to as ``full-polaron" if the variational parameters $\{f_{k,i}\}$ are simply set to $\{t_{k,i}\}$, the original system-reservoir couplings. 
If, instead, the optimal values for $\{f_{k,i}\}$ are obtained by minimizing the Gibbs-Bogoliubov-Feynman upper bound on the free energy, the transformation is called ``variational" \cite{Cao12,Cao_2016}. After the transformation, the Hamiltonian reads
\begin{equation}
    \hat{H}^\text{pol} = E_0\hat{I}+\hat{H}^\text{pol}_S+\hat{H}^\text{pol}_B+\hat{H}^\text{pol}_I
\end{equation}
where the energy shift is $E_0=\sum_k[f_{k,1}(f_{k,1}-2t_{k,1})/\nu_k+f_{k,2}(f_{k,2}-2t_{k,2})/\nu_k]$ with
\begin{equation}
    \hat{H}^\text{pol}_S = \Delta_{ 1}\kappa_1\hat{\sigma}^z_1+\Delta_{ 2}\kappa_2\hat{\sigma}^z_2-2E_I\hat{\sigma}^x_1\hat{\sigma}^x_2.
\end{equation}
Here, 
\begin{equation}
\kappa_i=\exp\left[-2\sum_k\frac{f^2_{k,i}}{\nu^2_k}\coth\left(\frac{\beta\nu_k}{2}\right)\right],
\end{equation}
and 
\begin{equation}
E_I = \sum_k[f_{k,1}(t_{k,2}-f_{k,2}/2)/\nu_k+f_{k,2}(t_{k,1}-f_{k,1}/2)/\nu_k].
\end{equation}
The bath Hamiltonian remains the same $\hat{H}^\text{pol}_B = \sum_k\nu_k\hat{c}^\dagger_k\hat{c}_k$, 
while the system-bath interaction Hamiltonian now reads
\begin{equation}
    \hat{H}^{\rm{pol}}_I = \sum^2_{i=1}\left(\hat{V}_{x,i},\hat{V}_{y,i},\hat{V}_{z,i}\right)\cdot\hat{\bf{\sigma}}_i
\end{equation}
where
\begin{equation}
\begin{aligned}
    \hat{V}_{x,i} =& \sum_k\left(t_{k,i}-f_{k,i}\right)\left(\hat{c}^\dagger_k+\hat{c}_k\right) \\
    \hat{V}_{y,i} =& -\Delta_i\sin\left[2{\rm{i}}\sum_k\frac{f_{k,i}}{\nu_k}\left(\hat{c}^\dagger_k-\hat{c}_k\right)\right]\\
    \hat{V}_{z,i} =&\Delta_i\cos\left[2{\rm{i}}\sum_k\frac{f_{k,i}}{\nu_k}\left(\hat{c}^\dagger_k-\hat{c}_k\right)\right]-\Delta_i\kappa_i
    \end{aligned}
\end{equation}
Here, $\hat{\sigma}_i$ is a vector of Pauli matrices. If we consider a full-polaron transform by setting $t_{k,i}=f_{k,i}$, and assuming that $\cos\left[2{\rm{i}}\sum_k\frac{f_{k,i}}{\nu_k}(\hat{c}^\dagger_k-\hat{c}_k)\right]$ is close to its average value of $\kappa_i$, then $\hat{V}_{y,i}$ is the only remaining interaction. With this mapping, we expect to observe similar dynamics as generated by the effective Hamiltonian in Fig.~\ref{fig:5}, Fig.~\ref{fig:6} and Fig.~\ref{fig:7}.


\section{Dynamics from the hierarchical equations of motion}
\label{ap:heom}

In general, in arriving at the Redfield equation Eq.~\eqref{eq:redfield_dyn} one relies on the Born-Markov approximation, which does not guarantee that the resulting equation leads to an appropriate definition of a generator of a dynamical semigroup. In turn, this may translate to non-positive dynamics of the reduced density matrix of the state. Though not typically an issue at weak system-reservoir coupling, non-positive dynamics may pose an issue in our calculations. Furthermore, the Redfield equation neglects non-Markovian effects, which may be important if the residual coupling is not weak.

In this Section, we however demonstrate that the dynamics generated by Eq.~\eqref{eq:redfield_dyn} is correct when compared to another numerically exact approach: the hierarchical equations of motion. We remark that our Markovian embedding introduced in Sec.~\ref{sec:effective} guarantees that the {\emph enlarged} system remains weakly coupled to the residual reservoir and, therefore, one would not expect positivity of the density matrix to be violated along the time trajectory generated by Eq.~\eqref{eq:redfield_dyn}. However, we find it important to confirm whether the correct dynamics are being generated in our model.

The hierarchical equations of motion allow one to study the dynamics of quantum systems coupled to reservoirs without recurring to weak-coupling perturbative analyses and/or the Born-Markov approximation~\cite{Tanimura:2020HEOM,Lambert:2023Qutip}. Although computationally more expensive than the techniques presented in this work, HEOM provides a reliable approach to benchmark our calculations.

We provide the basic concepts of HEOM here and refer the reader to Ref.~\cite{Lambert:2023Qutip} for further details. The method starts by discretizing the environment composed of a continuum of states and creating a hierarchy of equations of motion for auxiliary density operators that we need to solve simultaneously. For the appropriate dynamics to be obtained, one must maintain a sufficient amount of auxiliary operators in the hierarchy. Furthermore, there is an assumption that thermal correlation functions of the reservoir can be represented via a sum of exponentials. For specific spectral functions, analytical expressions can be obtained for these correlation functions and the sum must be truncated to a finite amount of exponential for the calculation to be numerically tractable.

\begin{figure}[t]
\fontsize{13}{10}\selectfont 
\centering
\includegraphics[width=0.91\columnwidth]{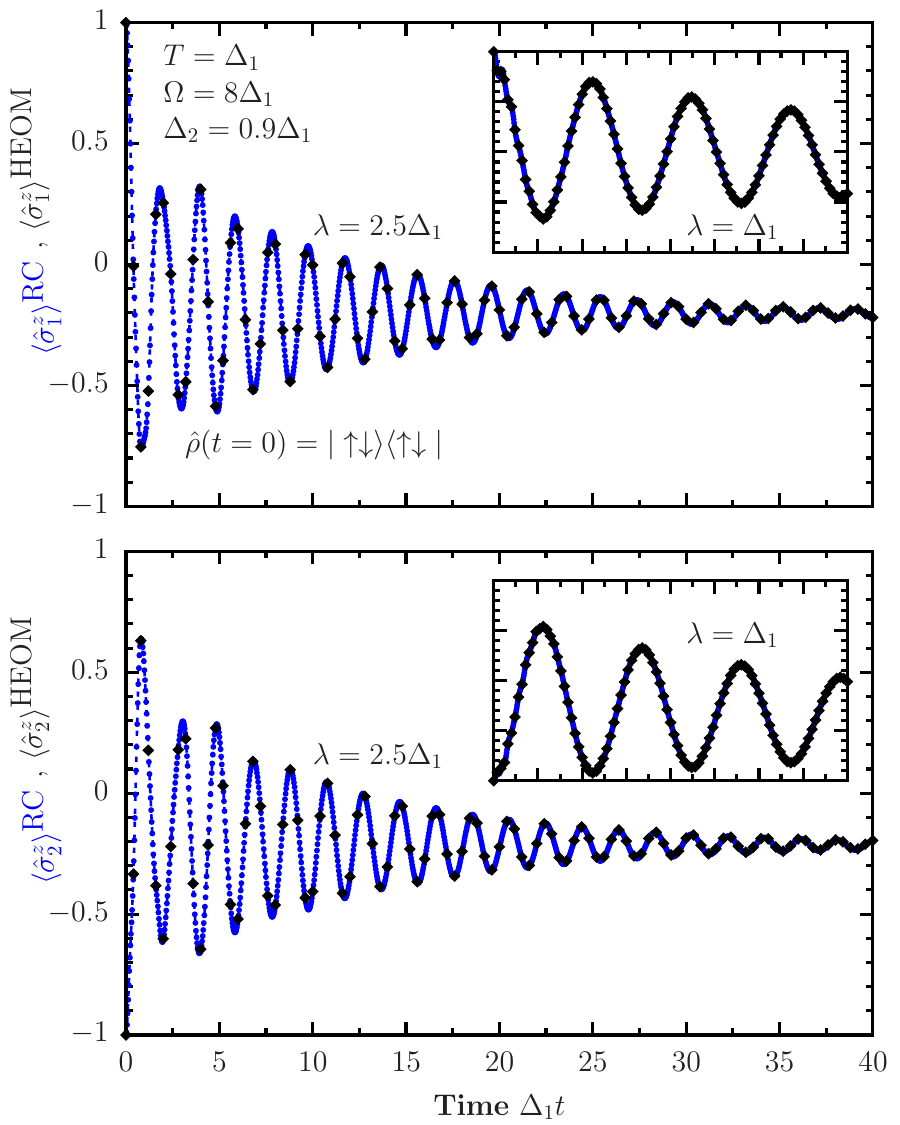}
\caption{Dynamics of the polarization in the two-spin system for first (top) and second (bottom) spins with a coupling parameter $\lambda = 2.5\Delta_1$. The reaction coordinate calculation was carried out via the Redfield Eq.~\eqref{eq:redfield_dyn} with $\gamma = 0.05$, $M = 25$, $\Delta_2 = 0.9\Delta_1$, $\Omega = 8\Delta_1$ and $T = \Delta_1$ (depicted as blue lines). 
The corresponding HEOM calculation (depicted as black diamonds) was done with parameters $N_c = 10$ and $N_k = 5$, which were sufficient to guarantee convergence in the time domain shown. Note that in the reaction-coordinate method, we use the Ohmic spectral function in Eq.~\eqref{eq:spec_fun_rc_ohmic} to characterize the residual bath, while the corresponding HEOM calculation uses the Brownian spectral density Eq.~\eqref{eq:spec_fun_brownian}, as it simulates the original bath.}
\label{fig:ap1}
\end{figure}

\begin{figure}[t]
\fontsize{13}{10}\selectfont 
\centering
\includegraphics[width=0.91\columnwidth]{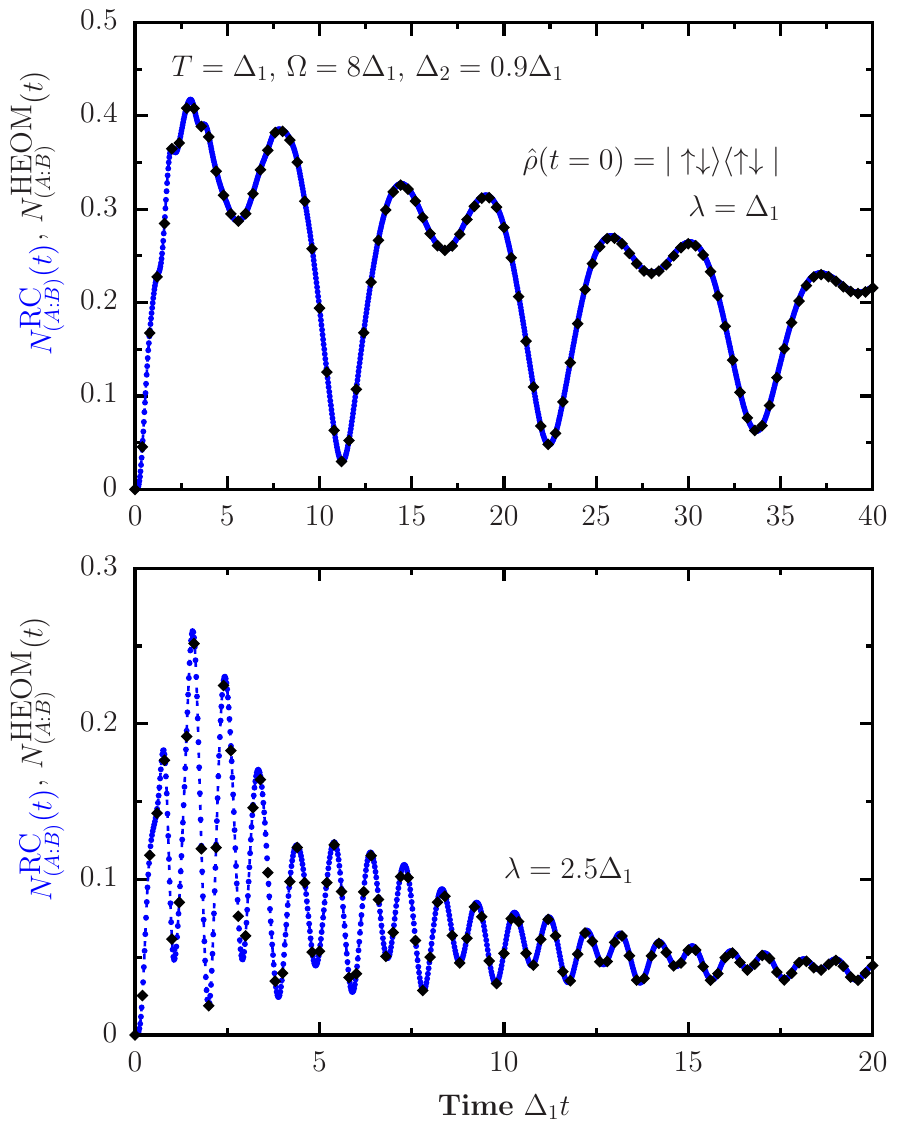}
\caption{Dynamics of the entanglement negativity in the two-spin system with $\lambda = \Delta_1$ (top) and $\lambda = 2.5\Delta_1$ (bottom). The reaction coordinate calculation was carried out via the Redfield Eq.~\eqref{eq:redfield_dyn} with $\gamma = 0.05$, $M = 25$, $\Delta_2 = 0.9\Delta_1$, $\Omega = 8\Delta_1$ and $T = \Delta_1$ (depicted as blue lines). 
The corresponding HEOM calculation (depicted as black diamonds) was done with parameters $N_c = 10$ and $N_k = 5$, which were sufficient to guarantee convergence in the time domain shown.
}
\label{fig:ap2}
\end{figure}

The bath correlation function is given by~\cite{Lambert:2023Qutip}
\begin{align}
C(t) &= \langle \hat{X}(\tau+t) \hat{X}(\tau) \rangle_B \\ \nonumber
&= \int_0^{\infty} {\rm{d}}\omega J(\omega)\left[ \coth \left(\frac{\beta\omega}{2} \right) \cos(\omega t) - \ii\sin(\omega t) \right ],
\end{align}
where $\hat X= \sum_{k} t_k (\hat c_k^{\dagger}+\hat c_k)$ (see Eq.~\eqref{eq:h_2spin_total}). The next step is to decompose the real and imaginary parts of $C(t)$ into a sum of exponentials
\begin{align}
C(t) = \sum_{k=1}^{N_R} c_k^R e^{-\gamma_k^R t} + \ii\sum_{k=1}^{N_I} c_k^I e^{-\gamma_k^I t},
\end{align}
where $N_R$ and $N_I$ are convergence parameters controlling the number of Matsubara terms in the real and imaginary parts of the expansion, respectively. In our calculations, we set $N_R = N_I = N_k$, where $N_k$ is the number of Matsubara terms. 
The expansion coefficients ($c_k^R$ and $c_k^I$) and the Matsubara frequencies ($\gamma_k^R$ and $\gamma_k^I$) can be real or complex, depending on the nature of the correlation functions of the bath and the spectral function being considered. In our work, we considered the Brownian spectral function for which these parameters can be found analytically (see Eqs.~(24)-(27) in Ref.~\cite{Lambert:2023Qutip}). 

The $n$-th equation in the hierarchy described above can be written as
\begin{align}
\label{eq:heom}
\dot{\rho}^n(t) &= \left( -\ii\hat{H}_S^{\times} - \sum_{j = R,I} \sum_{k=1}^{N_j} n_{jk} \gamma_k^j \right)\rho^n(t) \\ \nonumber
&- \ii\sum_{k=1}^{N_R} c_k^Rn_{Rk}\hat{S}^{\times} \rho^{n^-_{Rk}}(t) + \sum_{k=1}^{N_I} c_k^In_{Ik}\hat{S}^{\circ} \rho^{n^-_{Ik}}(t) \\ \nonumber
&- \ii\sum_{j = R,I} \sum_{k=1}^{N_j} \hat{S}^{\times}\rho^{n^+_{jk}}(t),
\end{align}
where $\hat{O}^{\times}\bullet = [\hat{O},\bullet]$ and $\hat{O}^{\circ}\bullet = \{\hat{O},\bullet\}$. Furthermore, $n = (n_{R1},n_{R2},...,n_{RN},n_{I1},n_{I2},...n_{IN})$ is a multidimensional index used to label the auxiliary density matrices with each $n_{jk}$ taking values in the set $\{0,1,2,\cdots,N_c\}$, with $N_c$ being a convergence parameter indicating the number of hierarchies to be included. We note that the states labeled with $n = (0,\cdots,0)$ correspond to the density operator of which we want to study the dynamics, while the $\rho^{n^{\pm}_{jk}}(t)$ correspond to auxiliary operators labeled with index $n_{jk}$ raised or lowered by one.

In our calculations, we employ the HEOM implementation found in the Quantum Toolbox in Python (QuTiP) package~\cite{QuTip1, QuTip2} to solve for the dynamics of the reduced density operator of the two-qubit system.
In Fig.~\ref{fig:ap1} we show that the reaction-coordinate approach to simulate the polarization dynamics (using the Redfield QME [Eq.~\eqref{eq:redfield_dyn}] for the RC Hamiltonian) excellently captures the dynamics compared to the HEOM calculation. In Fig.~\ref{fig:ap2} we display the corresponding dynamics for the entanglement negativity Eq.~\eqref{eq:neg_2}, showing yet again excellent agreement between the Redfield QME on the reaction-coordinate approach and the HEOM.

Though, in general, the Redfield equation does not preserve complete positivity of the reduced density operator along time trajectories, and it ignores non-Markovian effects, we find that in our case the true dynamics are indeed captured for the reaction coordinate Hamiltonian, for our particular system and bath configuration.

\bibliography{blibliography}

\begin{thebibliography}{98}
\bibitem [{Lam()}]{Lamb}%
  \BibitemOpen
  \href@noop {} {}\bibinfo {note} {Note that the Lamb shift generates coherent
  coupling, between e.g., the $\ket{\uparrow\downarrow}$ and the opposite
  excitation, $\ket{\downarrow\uparrow}$. However, this coupling is
  significantly smaller than the dissipative decay rate. Simulations were
  performed in \cite{Ivander}. There, in Fig. 11 we showed that for the
  relevant case (with spin splitting difference at the range of 10\% of the
  total gap, $\Delta =0.1$ in the notation of the above cited paper), the Lamb
  shift is two orders of magnitude smaller than the relaxation rate. At weak
  coupling, we thus do not expect the Lamb shift to generate synchronization,
  since the timescale is longer than the local relaxation
  timescale.}\BibitemShut {Stop}%
\end{thebibliography}

\end{document}